\documentclass[12pt, draftclsnofoot, onecolumn]{IEEEtran}
\makeatletter
\def\ps@headings{%
\def\@oddhead{\mbox{}\scriptsize\rightmark \hfil \thepage}%
\def\@evenhead{\scriptsize\thepage \hfil \leftmark\mbox{}}%
\def\@oddfoot{}%
\def\@evenfoot{}}
\makeatother 
\usepackage{amsfonts,dsfont}
\usepackage[dvips]{graphicx}
\usepackage{caption}
\usepackage{subfigure}
\usepackage{times}
\usepackage{cite}
\usepackage{lettrine}
\usepackage{amsmath}
\usepackage{bm}
\allowdisplaybreaks[4]
\usepackage{array}
\usepackage{amssymb}

\usepackage{stfloats}
\usepackage{slashbox}
\usepackage{graphicx}
\usepackage{footnote}
\usepackage{booktabs}
\usepackage{array}
\usepackage{algorithmic}
\usepackage{algorithm}
\usepackage{subeqnarray}
\usepackage{cases}
\usepackage{threeparttable}
\usepackage{color}
\usepackage{url}
\usepackage{enumerate}

\usepackage{cleveref} 

\makeatletter       
\renewcommand{\maketag@@@}[1]{\hbox{\m@th\normalsize\normalfont#1}}%
\makeatother
\makeatletter
\renewcommand{\fnum@figure}{Fig. \thefigure}
\makeatother

\DeclareMathOperator*{\argmax}{argmax}
\DeclareMathOperator*{\argmin}{argmin}

\newtheorem{theorem}{\underline{Theorem}}
\newtheorem{lemma}{\underline{Lemma}}

\newtheorem{proposition}{\underline{Proposition}}
\newtheorem{remark}{\underline{Remark}}

\begin{document}
\title{A Two-Timescale Approach to Mobility Management for Multi-Cell Mobile Edge Computing}
\author{Zezu Liang, Yuan Liu, Tat-Ming Lok, and Kaibin Huang

\thanks{Z. Liang and T. M. Lok are with Department of Information Engineering, The Chinese University of Hong Kong, Hong Kong (e-mail: zezuliang@gmail.com;  tmlok@ie.cuhk.edu.hk). Y. Liu is with school of Electronic and Information Engineering,
South China University of Technology, Guangzhou 510641, China (e-mail: eeyliu@scut.edu.cn). K. Huang is with Department of Electrical and Electronic Engineering, The University of Hong Kong, Hong Kong (e-mail: huangkb@eee.hku.hk).}
}

\maketitle

\vspace{-1.2cm}
\begin{abstract}
Mobile edge computing (MEC) is a promising technology for enhancing the computation capacities and features of mobile users by offloading complex computation tasks to the edge servers. However, mobility poses great challenges on delivering reliable MEC service required for latency-critical applications. First, mobility management has to tackle the dynamics of both user's location changes and task arrivals that vary in different timescales. Second, user mobility could induce service migration, leading to reliability loss due to the migration delay. In this paper, we propose a two-timescale mobility management framework by joint control of service migration and transmission power to address the above challenges. Specifically, the service migration operates at a large timescale to support user mobility in the multi-cell network, while the power control is performed at a small timescale for real-time task offloading. Their joint control is formulated as an optimization problem aiming at the long-term mobile energy minimization subject to the reliability requirement of computation offloading. To solve the problem, we propose a Lyapunov-based framework to decompose the problem into different timescales, based on which a low-complexity two-timescale online algorithm is developed by exploiting the problem structure. The proposed online algorithm is shown to be asymptotically optimal via theoretical analysis, and is further developed to accommodate the multiuser management. The simulation results demonstrate that our proposed algorithm can significantly improve the energy and reliability performance.
\end{abstract}


\section{Introduction}
The rapid development of advanced mobile applications and Internet-of-Things (IoT) calls for high quality of service (QoS), such as ultra-low latency, ultra-high reliability, robust security, enhanced broadband access, and ubiquitous connectivity. It is commonly agreed that these strict requirements cannot be fulfilled by the conventional cloud computing as central cloud is far from real-time data generated by edge users. Mobile (or multi-access) edge computing (MEC) has been proposed as a solution to address the issue by deploying cloud computing functions at the network edges \cite{whitepaper,MECsurvey1,MECsurvey2,ourwork,MEC3}. Specifically, MEC allows users to offload computation tasks to proximate edge servers [e.g., base stations (BSs) or access points] for execution. This avoids data transportation across backhaul networks and thereby reduces latency and traffic congestion. Given the dense geographical distribution of servers, MEC is envisioned as a promising platform for enabling the emerging computation-intensive and latency-critical applications, such as real-time online gaming and autonomous driving \cite{MECsurvey2}. In this paper, we investigate the mobility management problem in MEC, aiming at supporting the MEC applications under the presence of user mobility.

\vspace{-0.1in}
\subsection{Related Works}
As mobile users may traverse different cells, one challenge faced by designing multi-cell MEC networks is mobility management to guarantee uninterrupted service \cite{ETSI018, 9119487}. The direct way to support mobility is service migration \cite{SM,SM1,ourwork2}, namely, continuously migrate the ongoing computing services of mobile users to their dynamically associated servers/BSs along the users' traveling paths. However, the uncertainty of user mobility makes the optimal  migration policies difficult to design. Three main approaches have been developed to address this issue. The first one is based on the prediction of the short-term user mobility and service latency to make more informed migration decisions \cite{predict1, predict2}. The second approach involves online migration decision making based on modeling the user movement as a Markovian process and applying the theory of Markov decision process (MDP) to optimize the decisions \cite{MDP1, MDP2,multiuser2,multiuser3}. The limitation of such an approach lies in its requirement of statistical information of user mobility, which is not always available in practice.

The last approach, which is closely related to this work, focuses on online migration design without a priori knowledge of future user mobility. Specifically, learning-driven migration schemes are proposed in \cite{EMM,dynamic2} based on multi-armed bandit theory, and in \cite{RL} using the deep reinforcement learning approach, in which the user copes with the lack of prior knowledge using the trial-and-error method. On the other hand, by utilizing the Lyapunov optimization technique, an online migration strategy is proposed in \cite{multiuser1} that balances the service latency, the incurred migration cost, and the long-term user movement. The theory is also applied in \cite{relatedwork} to develop a framework of dynamic user-BS association to satisfy the application requirements of latency and reliability under constraints on the task queue lengths.

In view of prior work, two issues have not been addressed. First, only service migration is insufficient for guaranteeing the QoS for many latency-critical applications. In general, latency-critical tasks generated by the application often arrive at a smaller timescale than the service migration that adapts mainly to user movement. For instance, the update time in industrial IoT applications is between $0.5\!\sim\!500$ ms \cite{updatetime}, while the service migration is performed less frequently at the timescale of seconds to minutes in practice \cite{mobilitytimescale}. Second, the migration process induces reliability loss due to the migration delay. The BS handover procedure and migration of user's application profiles require a certain amount of time to complete (15 ms delay in a 5G handover scenario \cite{handoverdelay} for example). This leads to service interruption when tasks arrive during the migration process. To address these two issues motivates this work.

\vspace{-0.1in}
\subsection{Our Contributions}
In this paper, we consider a user moving in the multi-cell MEC network and aim to guarantee the reliability of the user's latency-critical application. We propose a novel mobility management framework that features the joint optimization of service migration and power control. Motivated by the fact that the user's location changes slower than the task arrivals, our proposed framework performs service migration at a large timescale to support user mobility and dynamic transmit power control at a small timescale to accommodate the real-time task offloading. In order to quantify the reliability loss caused by the mobile environment as well as the migration delay, we define the event of task failure as one that the offloading time exceeds the latency requirement or a task arrives during the migration process. The mentioned joint optimization aims at minimizing the long-term user's energy consumption while ensuring the reliability requirement that the probability of task failure is below a pre-defined threshold.

The main contributions of this paper are summarized as follows.
\begin{itemize}
\item We propose an online two-timescale control algorithm to solve the formulated problem. By invoking the Lyapunov optimization framework, our proposed algorithm can decouple the original two-timescale joint problem into two subproblems with different timescales, i.e., the service migration subproblems over the large timescale, and the power allocation subproblems over the small timescale. As the core components of the algorithm, we further derive the optimal migration policy and the optimal power strategy for solving these two subproblems, which allows to make online decisions in low complexity and without requiring any future information.

\item We prove that the proposed algorithm can achieve asymptotically optimal performance. Furthermore, the optimal power control is proved to have a threshold-based structure. Specifically, in each slot, a task is offloaded with the minimum required power if the power is below the threshold, and the task is dropped otherwise. In addition, it is shown that in each frame the user always migrates its service to the BS with the minimum weighted sum of energy consumption and task-dropping cost.

\item We also extend the online algorithm to multiuser management by designing an efficient per-frame migration scheme. The proposed scheme takes into account the load-balance factor in multiuser migration decisions, and it is based on the adjustment of the worst user-BS association to find a near-optimal solution.
\end{itemize}

The rest of this paper is organized as follows. We introduce the system model and formulate the problem in Section II and Section III, respectively. We design the online algorithm framework in Section IV, and provide the algorithm implementation and performance analysis in \makebox{Section V}. The extension to multiuser management is discussed in Section VI. Simulation results are presented in Section VII, and in Section VIII, we conclude the paper.

\section{System Model}
\begin{figure}[t]
\setlength{\abovecaptionskip}{0.3cm}
\setlength{\belowcaptionskip}{-0.4cm}
\begin{centering}
\includegraphics[width=0.5\linewidth]{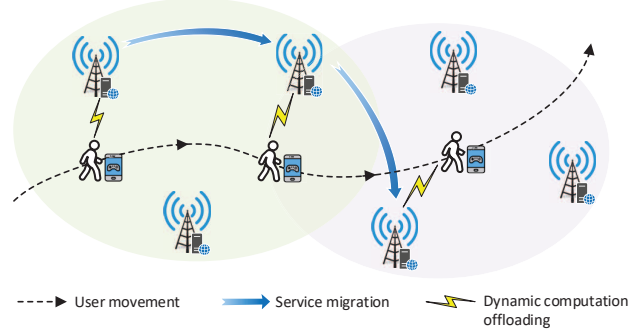}
\caption{System Model}
\label{fig:1}
\end{centering}
\end{figure}

As shown in Fig. \ref{fig:1}, we consider that a mobile user moves in a multi-cell network deployed with $N$ based stations (BSs), denoted by set $\mathcal{N}= \{1, 2, \cdots, N\}$. The network operates in a time-slotted manner, in which each time slot $t\in \{0, 1, \cdots\}$ has slot length $\tau$ that is consistent with the coherence time of the wireless channel. Each BS is integrated with an MEC server and can provide computing service. The application of the mobile user is computation-intensive such that all the generated tasks have to be offloaded to the BS (server) for execution. We assume that the computation tasks are homogeneous \cite{taskmodel1,taskmodel2,taskmodel3} and described by $A (L, \xi, \tau_d)$, in which $L$ (in bits) denotes the input data size of the task, $\xi$ denotes the number of CPU cycles required for processing the $L$-bit input data, and $\tau_d$ denotes the task latency requirement.  We consider a latency-critical scenario where the task latency requirement does not exceed the slot length, i.e., $\tau_d\leq \tau$. The task arrivals across slots are modeled as a Bernoulli process \cite{taskmodel1,taskmodel4,taskmodel5}. Specifically, at the beginning of each time slot $t$, a computation task $A(L, \xi, \tau_d)$ arrives with probability $\rho$, and with probability $1-\rho$, there is no task arrival. Therefore, let $a(t)\in \{0,1\}$ be the task arrival indicator. We have $\Pr(a(t)=1)=\rho$ and $\Pr(a(t)=0)=1-\rho$.

In order to satisfy the application's latency requirement, joint service migration and transmit power control are considered during the user movement. We assume that service migration occurs when the user changes its BS association from one to another and is conducted by joint communication handover and computation migration between the two BSs \cite{ETSI018,predict2,EMM,MAB2,multiuser1,ourwork2}. Here, computation migration refers to the migration of the user's application instances (or application state)\cite{ETSI018}. Meanwhile, the user performs dynamic transmit-power control for computation offloading as its serving BS and channel change. The corresponding models and assumptions are elaborated as follows.

\subsubsection{Two-Timescale Operation Model}
\begin{figure}[t]
\setlength{\abovecaptionskip}{0.3cm}
\setlength{\belowcaptionskip}{-0.4cm}
\begin{centering}
\includegraphics[width=0.8\linewidth]{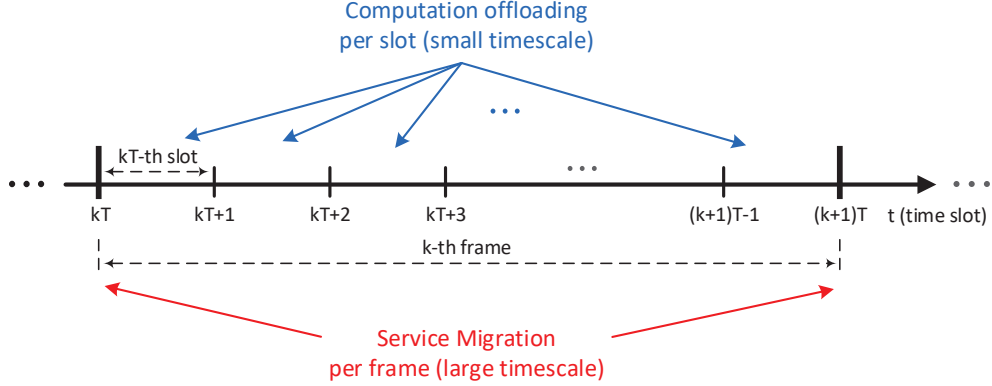}
\caption{The two-timescale model of service migration and computation offloading}
\label{fig:2}
\end{centering}
\end{figure}

Note that for the case of latency-critical applications, task offloading is often performed more frequently than service migration, due to the different timescales between the task arrivals and the user's location changes. For instance, the tasks generated from road safety are at the timescale of hundred milliseconds \cite{updatetime}, while the service migration occurs over the timescale of seconds to minutes \cite{mobilitytimescale} since it reacts mainly to the user movement and requires high operational cost. In this regard, we propose a two-timescale mobility management framework for large-timescale service migration and small-timescale power control as shown in Fig. \ref{fig:2}. Specifically, we group every consecutive $T$ time slots as a time frame, indexed by $k \in \{0, 1, \cdots\}$, and denote the set of time slots in the $k$-th frame as $\mathcal{T}_k = \{kT, kT+1, \cdots, (k+1)T-1\}$. We assume that:
\begin{itemize}
\item Large timescale: Service migration is made at the beginning of each frame and remains unchanged during a frame.
\item Small timescale: Transmit power control is performed at each time slot for task offloading.
\end{itemize}

\subsubsection{Service Migration Model}
At time slot $t=kT$, i.e., the beginning of a frame, the user determines the migration/association decision for the $k$-th frame. Let $n(k)\in \mathcal{N}$ denote the user's associated BS. Clearly, service migration is triggered when $n(k)\neq n(k-1)$. We assume that the migration operation can cause $C$ slots of service interruption (i.e., service migration delay) at the beginning of a frame, with $0\leq C <T$, during which computation offloading is temporarily disrupted. We further denote the set of time slots in frame $k$ for doing migration as $\mathcal{T}_{k}^{c}$. By definition, we have $\mathcal{T}_{k}^{c} = \{kT, \cdots, kT+C-1\}$ if $n(k)\neq n(k-1)$, and $\mathcal{T}_{k}^{c} = \{\emptyset\}$ otherwise.

\subsubsection{Computation Offloading Model}
At each time slot $t$, if a task arrives, the user adjusts the transmit power $p(t)$ based on the real-time channel condition to support task offloading. We denote the uplink channel power gain from the user to BS $n$ at slot $t$ as $h_n(t)$, which includes path loss (that captures the user's location change) and small-scale fading. We assume that $h_n(t)$ experiences block fading, i.e., $h_n(t)$ remains constant within each time slot but possibly varies over different time slots. Let $f_n(k)$ denote the computation rate (CPU cycles per second) of BS $n$ assigned to the user.  We assume that the BS adjusts $f_n(k)$ on a frame basis, since dynamic provisioning of the compute resource (e.g., virtual machines or containers) is often carried out at a larger time interval. Then, given the associated BS $n(k)=n$ and transmit power $p(t)$, the total latency for offloading and computing task $A(L, \xi, \tau_d)$ at slot $t \in \mathcal{T}_k$ can be expressed as
\begin{align}\label{eqn:D}
D(n, p(t))= \frac{L}{W\log_2\big(1+\frac{p(t)h_{n}(t)}{\sigma}\big)}+\frac{\xi}{f_{n}(k)},
\end{align}
where $W$ is the channel bandwidth and is assumed to be homogenous among BSs for simplicity, and $\sigma$ denotes the noise power.  We consider the task offloading and ignore the result downloading phase because of the relative much smaller sizes of computed results.

Accordingly, the user's energy consumption for offloading a task at slot $t \in \mathcal{T}_k$ is given by
\begin{align}
&E(n, p(t))= \frac{L p(t)}{W\log_2\big(1+\frac{p(t)h_{n}(t)}{\sigma}\big)}. \label{eqn:E}
\end{align}

\section{Problem Formulation}
Based on the proposed two-timescale mobility management framework, our goal is to design an online service migration and power control algorithm that minimizes the user's energy consumption and meanwhile satisfies the tasks' latency requirements continuously.
Nevertheless, due to the service migration delay and the wireless channel fluctuation, some of the arrived computation tasks may not be accomplished within the deadline, leading to task failure. For example, task failure may occur when the user's service is being migrated, or when the wireless channel from user to its associated BS is in a deep fade.
To take this aspect into consideration, we denote $X(t)\in\{0, 1\}$ as the task failure indicator, with $X(t)=1$ indicating the task failure occurs at slot $t$, and $X(t)=0$ otherwise. Then, given the BS association decision $n(k)=n$ and transmit power $p(t)$, the task failure event at slot $t\in \mathcal{T}_k$ can be characterized by
\begin{align}\label{eqn:1}
X(t)=\begin{cases}
\mathds{1}_{\left\{a(t)=1\right\}}, &\text{if~} t\in \mathcal{T}_{k}^{c}, \\
\mathds{1}_{\left\{a(t)=1, ~D\left(n, p(t)\right) >\tau_d \right\}}, &\text{if~} t\in \mathcal{T}_k\backslash\mathcal{T}_{k}^{c},
\end{cases}
\end{align}
where $\mathds{1}_{\{x\}}$ is the indicator function, with $\mathds{1}_{\{x\}} = 1$ if event $x$ is true and  $\mathds{1}_{\{x\}} = 0$ otherwise. \eqref{eqn:1} specifies that task failure occurs if there is a task arrival during the service migration process (i.e., $t\in \mathcal{T}_{k}^{c}$), or the arrived task can not be completed within the latency requirement.

The task failure events can degrade the service reliability for latency-critical applications. In this regard, we impose the following constraint on the average occurrence rate of task failure:
\begin{align}\label{eqn:reliable}
\lim_{K\rightarrow \infty} \frac{1}{KT}\sum_{k=0}^{K-1}\sum_{t\in\mathcal{T}_k} \mathbb{E}\left\{X(t)\right\}\leq \epsilon,
\end{align}
where $\epsilon \ll 1$ is the maximum threshold of the task-failure rate, which can be seen as the application's reliability requirement. The expectation $\mathbb{E}\{\cdot\}$ is taken over all sources of randomness, including task arrivals and dynamics of channel conditions and BSs' computation rates.

Similarly, combining the factors of task arrival and service interruption during migration, we can express the user's energy consumption at every slot $t\in \mathcal{T}_k$ as
\begin{align}\label{eqn:2}
\mathcal{E}(t)=\begin{cases}
0, &\text{if~}   t\in \mathcal{T}_{k}^{c}, \\
E(n, p(t))\cdot\mathds{1}_{\left\{a(t)=1, ~D\left(n, p(t)\right) \leq \tau_d \right\}}, &\text{if~}  t\in \mathcal{T}_k\backslash\mathcal{T}_{k}^{c},
\end{cases}
\end{align}
i.e., the user consumes energy only in the case when the arrived task can be accomplished within the latency requirement.

Incorporating the constraint \eqref{eqn:reliable}, our studied problem is to minimize the user's long-term energy consumption while ensuring the reliability requirement for the latency-critical application, which can be formulated as
\begin{subequations}
\begin{align}
(\text{P1})~~\min_{\{n(k)\},\{p(t)\}} \quad & \mathit{E_{\rm{av}}}\triangleq\lim_{K\rightarrow \infty} \frac{1}{KT}\sum_{k=0}^{K-1}\sum_{t\in\mathcal{T}_k} \mathbb{E}\left\{\mathcal{E}(t)\right\} \label{eqn:ob}\\
\rm{s.t.} \quad~~ ~~&\mathit{X_{\rm{av}}}\triangleq\lim_{K\rightarrow \infty} \frac{1}{KT}\sum_{k=0}^{K-1}\sum_{t\in\mathcal{T}_k} \mathbb{E}\left\{X(t)\right\}\leq \epsilon  \quad \quad \quad\label{eqn:st1}\\
&~n(k)\in \mathcal{N}, \quad \quad k=0, 1,  \cdots \quad \quad \quad \quad \quad\label{eqn:st2}\\
&~0\leq p(t)\leq \overline{P}, \quad  t=0, 1,  \cdots \label{eqn:st3}
\end{align}
\end{subequations}
where \eqref{eqn:st3} is the peak power constraint of the user.

There are two major challenges in solving Problem (P1). First, optimally solving \makebox{Problem (P1)} requires the complete information of the user trajectory, task arrivals, and network-level conditions over the entire time horizon, which is extremely difficult to acquire in advance. Second, the migration decision $n(k)$ and the power allocation $p(t)$ that change in different timescales, are tightly coupled, e.g., the migration decision $n(k)$ for the $k$-th frame affects the power allocations $\{p(t)\}$ in slots $t\in\mathcal{T}_k$, and vice versa. To address the above challenges, we develop an online two-timescale control algorithm in the following two sections.

\section{Online Two-Timescale Algorithm Design}
In this section, we present the framework design of our online algorithm. First, we transform Problem (P1) into an online optimization problem using the Lyapunov technique. Subsequently, a two-timescale control algorithm is designed to solve the transformed problem optimally.

\vspace{-0.1in}
\subsection{Problem Transformation}
In order to take the advantage of Lyapunov optimization, we first convert the reliability constraint \eqref{eqn:st1} into an equivalent queue stability constraint, which is described as follows. We construct a virtual queue with the queue length evolving according to $X(t)$ and $\epsilon$ as
\begin{align}\label{eqn:virtualqueue}
Q(t+1) = \left[Q(t)+X(t)-\epsilon\right]^+, \quad t=0, 1, \cdots
\end{align}
where $[\cdot]^+\triangleq \max\{\cdot, 0\}$. $Q(t)$ is the queue length at slot $t$, with $Q(0)=0$, which indicates how far the current task-failure backlog exceeds the threshold $\epsilon$. According to the Lyapunov optimization theory \cite{Lyapunov1}, the long-term time-averaged constraint \eqref{eqn:st1} is equivalent to the mean-rate stability constraint on the virtual queue, i.e., $\lim_{t\rightarrow\infty}\mathbb{E}\{Q(t)\}/t \rightarrow 0$.

To proceed, we define a $T$-slot (i.e., frame-based) conditional Lyapunov drift as
\begin{align}\label{eqn:deltaT}
\Delta_T(Q(t))\triangleq \mathbb{E}\bigg\{\frac{1}{2}Q(t+T)^2 - \frac{1}{2}Q(t)^2\Big|Q(t)\bigg\}.
\end{align}
Given the current queue length $Q(t)$, $\Delta_T(Q(t))$ characterizes the expected change in quadratic function of the queue length after $T$ time slots. Intuitively, minimizing $\Delta_T(Q(t))$ in each $T$ slots can prevent the queue length from unbounded growth and thus stabilize the queue.

Recalling that our problem objective is to minimize the energy consumption defined in \eqref{eqn:ob}, we add the energy consumption (as a penalty function) into \eqref{eqn:deltaT} to obtain the following drift-plus-penalty term for the $k$-th frame:
\begin{align}
\mathcal{D}(Q(kT)) \triangleq\Delta_T(Q(kT)) +V\mathbb{E}\bigg\{\sum_{t\in \mathcal{T}_k}\mathcal{E}(t)\Big|Q(kT)\bigg\}.
\end{align}
where $V \geq0$ is a control parameter, indicating an importance weight on how much we emphasize the energy consumption minimization.

The main idea of the Lyapunov optimization-based algorithm is to minimize the upper bound of the drift-plus-penalty term $\mathcal{D}(Q(kT))$ for joint queue stability and energy consumption minimization. To this end, we have the following two lemmas regarding the upper bound of $\mathcal{D}(Q(kT))$ for our two-timescale algorithm design.
\begin{lemma}\label{lem1}
Under any feasible decisions $n(k)\in \mathcal{N}$ and $0\leq p(t) \leq \overline{P}, \forall t\in \mathcal{T}_k$, $\mathcal{D}(Q(kT))$ is upper bounded by
\begin{align}\label{eqn:lemma1}
\mathcal{D}(Q(kT))\leq& B_1T + V\mathbb{E}\bigg\{\sum_{t\in \mathcal{T}_k}\mathcal{E}(t)\Big|Q(kT)\bigg\}+\mathbb{E}\bigg\{\sum_{t\in\mathcal{T}_k}Q(t)\left[X(t)-\epsilon\right] \Big|Q(kT)\bigg\}.
\end{align}
Here, $B_1\triangleq \frac{1}{2}(\rho+\epsilon^2)$ is a constant.
\end{lemma}
\begin{IEEEproof}
See Appendix A.
\end{IEEEproof}

The upper bound given in Lemma \ref{lem1} [i.e., the R.H.S. of \eqref{eqn:lemma1}] is widely used in the single-timescale control problems \cite{multiuser1} (i.e., frame size $T=1$). However, it is difficult to be applied directly to the two-timescale case since minimizing the R.H.S. of \eqref{eqn:lemma1} at the beginning of every frame $t=kT$ requires the future information of $\{Q(t)\}$ over $t\in [kT+1, ..., (k+1)T-1]$, which is hard to be predicted in practice due to its accumulative nature over time slots. To address this issue, we further relax the R.H.S. of \eqref{eqn:lemma1} as shown in the following lemma \cite{Lyapunov1,Lyapunov2,Lyapunov3}.
\begin{lemma}\label{lem2}
Under any feasible decisions $n(k)\in \mathcal{N}$ and $0\leq p(t) \leq \overline{P}, \forall t\in \mathcal{T}_k$, we have
\begin{align}\label{eqn:lemma2}
\mathcal{D}(Q(kT))\leq& B_2T + \mathbb{E}\bigg\{\sum_{t\in \mathcal{T}_k}V\mathcal{E}(t)+Q(kT)\left[X(t)-\epsilon\right]\Big|Q(kT)\bigg\}.
\end{align}
Here, $B_2\triangleq B_1+(T-1)[(1-\epsilon)\rho+\epsilon^2]/2$ is a constant.
\end{lemma}
\begin{IEEEproof}
See Appendix B.
\end{IEEEproof}

The upper bound in Lemma \ref{lem2} is derived from the R.H.S. of \eqref{eqn:lemma1} by approximating the future queue length values as the current value at slot $kT$, i.e., $Q(t)\approx Q(kT)$ for all $t\in [kT+1, \cdots, (k+1)T-1]$. This approximation avoids the prediction of future queue lengths, which significantly reduces the complexity and suits more on the two-timescale design. Furthermore, as will be proved in Section V-D, this approximation preserves the asymptotic optimality of our proposed algorithm.

\vspace{-0.1in}
\subsection{Algorithm Design}
We now present the online two-timescale algorithm design. The idea of the algorithm is to minimize the drift-plus-penalty upper bound in \eqref{eqn:lemma2} (i.e., the second term on the R.H.S.), subject to the constraints \eqref{eqn:st2} and \eqref{eqn:st3}, which can be proved to achieve a good performance for the original Problem (P1).  Specifically, our algorithm works in an online manner and takes the following three control actions:
\begin{itemize}
\item (\textbf{Migration decision per frame}) At time slot $t=kT$, with $k=0, 1, \cdots$,  the user observes $Q(kT)$, $n(k-1)$ and $f_n(k), \forall n$, and decides the optimal BS association $n^*(k)$ by solving the following per-frame problem:
        \begin{align}\label{eqn:perframeproblem}
        \min_{n(k), \{p(t)\}}\quad &\mathbb{E}\bigg\{\sum_{t\in \mathcal{T}_k}V\mathcal{E}(t)+Q(kT)X(t)\bigg\}\\
        {\rm{s.t.}}~~ \quad& n(k)\in\mathcal{N}, ~\quad 0\leq p(t)\leq \overline{P}, ~\forall t\in \mathcal{T}_k, \nonumber
        \end{align}
        The expectation $\mathbb{E}\{\cdot\}$ here is taken over the task arrival $a(t)$ and the channel randomness $\{h_n(t), \forall n\}$, for all $t\in \mathcal{T}_k$.

    \item (\textbf{Power allocation per slot}) At every slot $t\in\mathcal{T}_k$, given the BS association $n(k)$, the user observes the real-time channel condition $h_{n(k)}(t)$ and task arrival $a(t)$, and decides the power allocation $p^*(t)$ by solving the following per-slot problem:
        \begin{align}\label{eqn:perslotproblem}
        \min_{0\leq p(t)\leq \overline{P}}\quad &V\mathcal{E}(t)+Q(kT)X(t).
        \end{align}
    \item (\textbf{Queue update}) At each slot $t\in\mathcal{T}_k$, based on the obtained $p^*(t)$, compute $X(t)$ \makebox{by \eqref{eqn:1}} and update the virtual queue $Q(t)$ according to \eqref{eqn:virtualqueue}.
\end{itemize}

We next develop the optimal solutions to the subproblems \eqref{eqn:perframeproblem} and \eqref{eqn:perslotproblem}, respectively, which are the two building blocks for the algorithm implementation.

\vspace{-0.05in}
\section{Algorithm Implementation and Performance Analysis} \label{Sec:5}
In this section, we derive the optimal power strategy and the optimal migration policy for solving the per-slot Problem \eqref{eqn:perslotproblem} and the per-frame Problem \eqref{eqn:perframeproblem}, respectively. We also discuss the optimal migration mechanism for some special cases and analyze the algorithm performance in the end.

\vspace{-0.1in}
\subsection{Real-Time Power Allocation}
For the per-slot power allocation Problem \eqref{eqn:perslotproblem}, first we can easily obtain that $p^*(t)=0$ in two cases: 1) when $t\in \mathcal{T}_{k}^{c}$, i.e., during the service migration slots; and 2) when $a(t)=0$, i.e., no task arrival at slot $t$.

For the residual case that tasks arrive at the offloadable slots,  i.e., $t\in \mathcal{T}_{k}\backslash\mathcal{T}_{k}^{c}$ with $a(t)=1$, we can rewrite the corresponding per-slot Problem \eqref{eqn:perslotproblem} conditioned on $n(k)=n$ as
\begin{align}\label{eqn:perslotproblem2}
z_n(t)\triangleq\min_{0\leq p(t)\leq \overline{P}}&VE(n, p(t))\cdot\mathds{1}_{\left\{D\left(n, p(t)\right) \leq \tau_d \right\}}+Q(kT)\cdot\mathds{1}_{\left\{D\left(n, p(t)\right) >\tau_d \right\}},
\end{align}
where \eqref{eqn:perslotproblem2} is derived from \eqref{eqn:perslotproblem} by expanding $\mathcal{E}(t)$ and $X(t)$ according to the definitions in \eqref{eqn:2} and \eqref{eqn:1}.

From \eqref{eqn:perslotproblem2}, we can observe that based on whether the latency requirement is met, the user can choose to consume $E(n, p(t))$ amount of energy to offload the arrived task, or choose not to offload at the expense of $Q(kT)$. The virtual queue length $Q(kT)$ here acts as the price of dropping a task. A higher $Q(kT)$ emphasizes more on reliability, i.e., the arrived tasks should be successfully offloaded as much as possible; while a lower $Q(kT)$ prefers energy saving and tolerates more task failures. Intuitively, through the queue evolution, the performance of energy consumption and task failure can adaptively be coordinated over frames.

Next, we specify the optimal power strategy for Problem \eqref{eqn:perslotproblem2} as follows.

\begin{proposition}[Optimal Power Strategy for Per-Slot Offloading]\label{pro1} The optimal transmit power for Problem  \eqref{eqn:perslotproblem2} is given by
\begin{align}\label{eqn:pt}
p^*(t)=\begin{cases}
p_{n}^{\rm{min}}(t), &\text{if ~}p_{n}^{\rm{min}}(t)\leq p_{n}^{\rm{max}}(k),\\
0, &\text{otherwise},
\end{cases}
\end{align}
where $p_{n}^{\rm{min}}(t)$ is the minimum transmit power at slot $t$ to meet the task latency requirement, while $p_{n}^{\rm{max}}(k)$ denotes the maximum power allowed for per-slot offloading during $k$-th frame, which are respectively defined as:
\begin{align}
p_{n}^{\rm{min}}(t) &\triangleq \frac{\sigma}{h_n(t)}\bigg(2^{\frac{L}{W\left[\tau_d-\frac{\xi }{f_n(k)}\right]^+}}-1\bigg), \label{eqn:pn_min}\\
p_{n}^{\rm{max}}(k)&\triangleq \min\left\{ \tfrac{Q(kT)}{V\left[\tau_d-\frac{\xi}{f_n(k)}\right]^+}, \overline{P}\right\}.\label{eqn:pnmax}
\end{align}
\end{proposition}
\begin{IEEEproof}
It can be checked from \eqref{eqn:D} and \eqref{eqn:E} that $D(n, p(t))$ is monotonically decreasing while $E(n, p(t))$ is monotonically increasing with $p(t)$, $\forall n\in \mathcal{N}$. By letting \makebox{$D(n, p(t))=\tau_d$}, we obtain $p_{n}^{\rm{min}}(t)$ in \eqref{eqn:pn_min} as the minimum required power for meeting  the latency constraint, and $p^*(t)=p_{n}^{\rm{min}}(t)$ to achieve the minimum energy consumption in each task offloading.

We also note that when $VE(n, p_{n}^{\rm{min}}(t))> Q(kT)$ in \eqref{eqn:perslotproblem2}, i.e., the minimum energy consumption required for task offloading is higher than the task-dropping price, the task should be dropped for saving energy, thus $p^*(t)=0$. Let $VE(n, p_{n}^{\rm{min}}(t))= Q(kT)$ and further incorporate the peak power constraint \eqref{eqn:st2}, we can obtain $p_{n}^{\rm{max}}(k)$ in \eqref{eqn:pnmax} and the condition $p_{n}^{\rm{min}}(t)\!<\!p_{n}^{\rm{max}}(k)$ in \eqref{eqn:pt}, which completes the proof.
\end{IEEEproof}

Proposition \ref{pro1} reveals that the optimal power strategy for Problem \eqref{eqn:perslotproblem2} follows a \emph{threshold-based policy}. When  $p_{n}^{\rm{min}}(t)$ is below the threshold $p_{n}^{\rm{max}}(k)$, the user offloads the arrived task in power $p_{n}^{\rm{min}}(t)$; otherwise, the user should drop the task (i.e., $p^*(t)=0$) to avoid excessive energy consumption. Notably, $p_{n}^{\rm{min}}(t)$ in \eqref{eqn:pn_min} changes over each slot, adapting to the real-time channel condition $h_n(t)$, while threshold $p_{n}^{\rm{max}}(k)$ in \eqref{eqn:pnmax} remains unchanged within a frame but it is adjusted from one frame to another according to the updated $Q(kT)$.

\vspace{-0.1in}
\subsection{Migration Decision Per Frame}
In this subsection, we find the optimal $n^*(k)$ by solving the \makebox{per-frame Problem \eqref{eqn:perframeproblem}}. Recall that Problem \eqref{eqn:perframeproblem} is an expectation minimization problem. In order to compute the expectation, we make assumptions \cite{Lyapunov2,Lyapunov3} that the channel randomness is independent and identically distributed (i.i.d.) over the slots of a frame, and that the user has the statistical knowledge of channels in the current frame (but not the future frames).

According to the optimal power strategy in Proposition \ref{pro1}, we can derive the expected optimal per-slot performance for \makebox{Problem \eqref{eqn:perslotproblem2}} as follows.
\begin{theorem}\label{pro2}
Suppose that $h_n(t)$ is i.i.d. over the slots of a frame with the probability density function (PDF) denoted by $f_{h_n}(k)$. Then, for all $t\in\mathcal{T}_k\backslash\mathcal{T}_k^c$, the expectation of $z_n(t)$ taken over the channel randomness $h_n(t)$ is obtained as
\begin{align}\label{eqn:Znk}
&\mathbb{E}\left\{z_n(t)\right\} =  Ve_n(k)\!\int_{h_{n}^{\rm{min}}(k)}^{\infty}\!\frac{1}{h}f_{h_n}(k)dh +Q(kT)\Pr[h_n(t)<h_{n}^{\rm{min}}(k)]  \triangleq Z_{n}(k),
\end{align}
where $\Pr[\cdot]$ is the probability function, $e_n(k)\triangleq \sigma[\tau_d-\frac{\xi }{f_n(k)}]^+\!\Big(2^{\frac{L}{W[\tau_d-\frac{\xi}{f_n(k)}]^+}}\!-\!1\Big)$, and
\begin{align}\label{eqn:hnmin}
h_{n}^{\rm{min}}(k)\triangleq \frac{\sigma}{p_{n}^{\rm{max}}(k)}\!\Big(2^{\frac{L}{W[\tau_d-\frac{\xi }{f_n(k)}]^+}}\!-\!1\Big)
\end{align}
is the minimum threshold of channel gain to launch task offloading. In other words, task dropping occurs when $h_n(t)<h_{n}^{\rm{min}}(k)$.
\end{theorem}
\begin{IEEEproof}
According to Proposition \ref{pro1} and by comparing $p_{n}^{\rm{min}}(t)$ with $p_{n}^{\rm{max}}(k)$, we can derive
\begin{align}
z_n(t)=
\begin{cases}
V p_{n}^{\rm{min}}(t) \left[\tau_d- \frac{\xi}{f_n(k)}\right]^+ = \frac{V e_n(k)}{h_n(t)},  & \text{if~} h_n(t)>h_{n}^{\rm{min}}(k),\\
Q(kT),  & \text{otherwise},
\end{cases}
\end{align}
for each slot $t\in \mathcal{T}_k \backslash \mathcal{T}_k^c$. Taking the expectation on $z_n(t)$ over the random variable $h_n(t)$, we can obtain $\mathbb{E}\left\{z_n(t)\right\}$ as in \eqref{eqn:Znk}. Since $h_n(t)$ follows the same distribution among slots $t\in\mathcal{T}_k$, $\mathbb{E}\left\{z_n(t)\right\}$'s are identical for all $t\in\mathcal{T}_k\backslash\mathcal{T}_k^c$, which completes the proof.
\end{IEEEproof}

Different from $z_n(t)$ in \eqref{eqn:perslotproblem2}, $Z_n(k)$ in \eqref{eqn:Znk} represents the minimum expected execution cost (i.e., weighted sum of energy consumption and task-dropping cost) for each slot $t\in \mathcal{T}_k \backslash \mathcal{T}_k^c$ with task arrival and under a stationary channel environment. Note that $e_n(k)$ and $h_{n}^{\rm{min}}(k)$ in \eqref{eqn:Znk} are known constants to the user, since $f_n(k)$ and $Q(kT)$ (that affects $p_{n}^{\rm{max}}(k)$) are known at the beginning of the $k$-th frame. Therefore, with the statistical knowledge of channels, the user is able to compute $Z_{n}(k)$ by \eqref{eqn:Znk} at the beginning of each frame $t=kT$.

We define $Z_n^{\emph{sum}}(k)$ as the optimal objective value of the per-frame Problem \eqref{eqn:perframeproblem} under the association $n(k)=n$,  i.e.,
\begin{align}\label{eqn:Znk_sum_def}
Z_n^{\emph{sum}}(k)\triangleq \min_{\substack{0\leq p(t)\leq \overline{P}\\[0.1mm] \forall t\in \mathcal{T}_k}} \mathbb{E}\bigg\{\sum_{t\in \mathcal{T}_k}V\mathcal{E}(t)+Q(kT)X(t)\Big|n(k)=n\bigg\}, \quad \forall n.
\end{align}

As $a(t)$ and $h_n(t)$ are i.i.d. over slots $t\in \mathcal{T}_k$, $Z_n^{\emph{sum}}(k)$ in \eqref{eqn:Znk_sum_def} can be decoupled into $T$ independent per-slot problems with expectation minimization, each solved by the optimal power strategies discussed in the last subsection. Hence, we can further express $Z_n^{\emph{sum}}(k)$ as follows (see Appendix C):
\begin{align}\label{eqn:Znk_sum}
Z_n^{\emph{sum}}(k)=\begin{cases}
\rho (T-C)Z_{n}(k) + \rho C Q(kT), &\text{if ~} n\neq n(k-1) \\
\rho T Z_{n}(k), &\text{if ~} n=n(k-1)
\end{cases}, \quad \forall n, k.
\end{align}
Then, the optimal migration decision $n^*(k)$ for the $k$-th frame can be obtained by
\begin{align}\label{eqn:Nk_opt}
n^*(k)= \arg\min_{n\in\mathcal{N}} \left\{Z_n^{\emph{sum}}(k)\right\}.
\end{align}
From \eqref{eqn:Znk_sum}, we can see that the migration operation causes an expected $\rho C$ amount of task failure, which is a constant independent of which BS the user chooses to migrate to. Thus, for the BS set $n\in \mathcal{N}\backslash n(k-1)$, we have $\arg\min_{n\in\mathcal{N}\backslash n(k-1)}\left\{Z_n^{\emph{sum}}(k)\right\}=\arg\min_{n\in\mathcal{N}\backslash n(k-1)}\left\{Z_n(k)\right\}$. Using this result, we can express the optimal migration decision \eqref{eqn:Nk_opt}, in the form of the following migration policy:
\begin{align}\label{eqn:associa_policy}
n^*(k)=\begin{cases}
n^\prime, &\text{if ~} (1-\alpha) Z_{n^\prime}(k)+ \alpha Q(kT)\leq Z_{n(k-1)}(k), \\
n(k-1), &\text{otherwise~},
\end{cases}
\end{align}
where $n^\prime\triangleq \arg\min_{n\in\mathcal{N}\backslash n(k-1)}\left\{Z_n(k)\right\}$ and $\alpha\triangleq \frac{C}{T}$, with $0\leq \alpha \leq 1$, denoting the ratio of migration delay to a frame length.

The policy \eqref{eqn:associa_policy} suggests that the user always chooses migrating to the BS with the smallest $Z_n(k)$ whenever it performs a service migration, and the migration occurs only if condition $(1-\alpha) Z_{n^\prime}(k)+ \alpha Q(kT)\leq Z_{n(k-1)}(k)$ is met.

By incorporating the above migration policy and power strategy into the algorithm framework, we summarize the proposed online algorithm in Algorithm \ref{alg:1}.
\begin{algorithm}[t]
\caption{Online Two-Timescale Algorithm}
\begin{algorithmic}[1]\label{alg:1}
\STATE Set $V\geq 0$, $\epsilon \in (0, 1)$, and $n(-1)$ as the user's current associated BS.
\STATE Initialize $t=0$ and $Q(0)=0$.
\FOR{each frame $k=0, 1, \cdots, K-1$}
    \STATE Compute $n(k)$ by \eqref{eqn:associa_policy}.
    \STATE Set $\mathcal{T}_k^c=\{kT, kT+1, \cdots, kT+C-1\}$ if $n(k)\neq n(k-1)$, and otherwise $\mathcal{T}_k^c=\{\emptyset\}$.
    \FOR{each slot $t=kT, kT+1, \cdots, kT+T-1$}
        \IF{$t\in \mathcal{T}_k\backslash \mathcal{T}_k^c$ and $a(t)=1$}
            \STATE Compute $p(t)$ by \eqref{eqn:pt}.
        \ELSE
            \STATE Set $p(t)=0$.
        \ENDIF
        \STATE Compute $X(t)$ and $\mathcal{E}(t)$ by \eqref{eqn:1} and \eqref{eqn:2}, respectively.
        \STATE Update $Q(t)$ by \eqref{eqn:virtualqueue}.
    \ENDFOR
\ENDFOR
\ENSURE $\{n(k)\}$ and $\{p(t)\}$.
\end{algorithmic}
\end{algorithm}

\vspace{-0.1in}
\subsection{Properties of Optimal Migration Policy}
In this subsection, we derive additional insights into the migration policy in \eqref{eqn:associa_policy} for a concrete channel model. Specifically, we assume the channel power gain $h_n(t)$, $\forall t\in\mathcal{T}_k$ and $\forall n\in \mathcal{N}$, can be represented by
\begin{align}\label{eqn:hnt}
h_n(t) = g_n(t) H_n(k),
\end{align}
where $g_n(t)$ accounts for the small-scale fading power component at slot $t$ and $H_n(k)$ represents the large-scale fading power component in the $k$-th frame. $g_n(t)$ is assumed to be i.i.d. unit mean exponential random variables, i.e., the Rayleigh fading model considered for the fast fading. $H_n(k)$ captures the path loss and shadowing whose changes matches the timescale of a frame.

Building on the above channel model, we show in the sequel that the migration policy \eqref{eqn:associa_policy} has more straightforward migration mechanism for several special cases.

\subsubsection{Homogenous Computation Rates} Consider the case of $f_n(k)=f(k)$, $\forall n$, where $f(k)>\frac{\xi}{\tau_d}$ for the edge-execution feasibility. Then, $\left\{e_n(k), h_n^{\rm{min}}(k)\right\}$ are identical for all $n$ (see \makebox{Theorem \ref{pro2}}), and can be re-notated by $\left\{e(k), h^{\rm{min}}(k)\right\}$. We show that in this case, the migration decision for the $k$-th frame can be determined by simply comparing the parameter $H_n(k)$ of each BS.
\begin{proposition}[Homogenous Computation Rates]\label{pro3}
Assume that $f_n(k)=f(k)>\frac{\xi}{\tau_d}$, for all $n$. The following properties hold:
\begin{enumerate}[a)]
\item $n^\prime=\argmax_{n\in\mathcal{N}\backslash n(k-1)}\left\{H_n(k)\right\}$, i.e., if the user needs a service migration, it will always choose migrating to the BS with the highest $H_n(k)$.
\item The user keeps associating with the serving BS $n(k-1)$ if $H_{n(k-1)}(k)$ satisfies
\begin{align}\label{eqn:Hth1}
H_{n(k-1)}(k)> \frac{h^{\rm{min}}(k)}{\ln\left(\frac{1}{1-\alpha}\right)} \triangleq h_\alpha(k).
\end{align}
\item When the condition \eqref{eqn:Hth1} becomes invalid, the user migrates from BS $n(k-1)$ to BS $n^\prime$ if
\begin{gather}
H_{n(k-1)}(k)\leq H_{n^\prime}(k) \cdot \min \left\{\frac{h_\alpha(k)}{h_\alpha(k)+ H_{n^\prime}(k)}, \frac{1}{2}\right\}. \label{eqn:Hth2}
\end{gather}
\end{enumerate}
\end{proposition}
\begin{IEEEproof}
See Appendix D.
\end{IEEEproof}
\begin{remark}[Migration Policy]
Proposition \ref{pro3} reveals straightforward migration policies for this case. First, in each frame the user always selects the nearest BS (i.e., the highest $H_n(k)$), if all the BSs have the same computing rate. Second, when the channel gain of the serving BS is above a threshold specified by \eqref{eqn:Hth1}, there is no need of migration. Finally, when the condition \eqref{eqn:Hth1} becomes invalid, a migration is triggered if the new association can obtain a sufficient channel enhancement as specified by \eqref{eqn:Hth2}.
\end{remark}

\subsubsection{Heterogenous Computation Rates} Here we consider heterogenous computation rates by assuming no peak power \makebox{constraint \eqref{eqn:st3}}. Note that the transmit power still is bounded by $p_n^{\max}(k)=\frac{Q(kT)}{V[\tau_d-\frac{\xi}{f_n(k)}]^+}$ in \eqref{eqn:pnmax}. We obtain  for this case the migration decision relies on two parameters $H_n(k)$ and $h_n^{\rm{min}}(k)$, in which  $h_n^{\rm{min}}(k)$ [see \eqref{eqn:hnmin}]  is monotonically decreasing with $f_n(k)$.
\begin{proposition}[Heterogenous Computation Rates]\label{pro4}
Assume that $\overline{P}=\infty$ \footnote{Proposition \ref{pro4} also holds for the finite peak power as long as $\overline{P}>\frac{Q(kT)}{V[\tau_d-\frac{\xi}{f_n(k)}]^+}$ is met, $\forall n$.} and $f_n(k)>\frac{\xi}{\tau_d}$, for all $n$. Let $\nu_n(k)\triangleq \frac{h_n^{\rm{min}}(k)}{H_n(k)}$, for all $n\in\mathcal{N}$.  The following properties hold:
\begin{enumerate}[a)]
\item $n^\prime=\argmin_{n\in\mathcal{N}\backslash n(k-1)}\left\{\nu_n(k)\right\}$.
\item The user keeps associating with the serving BS $n(k-1)$ if $\nu_n(k)$ itself satisfies
\begin{align}\label{eqn:nu1}
\nu_n(k)<\ln\left(\tfrac{1}{1-\alpha}\right).
\end{align}
\item The user migrates the association from BS $n(k-1)$ to BS $n^\prime$ if
\begin{gather}\label{eqn:nu2}
\nu_{n(k-1)}(k)\geq \max\left\{\nu_{n^\prime}(k)+\ln\left(\tfrac{1}{1-\alpha}\right),~ 2\nu_{n^\prime}(k)\right\}.
\end{gather}
\end{enumerate}
\end{proposition}
\begin{IEEEproof}
See Appendix E.
\end{IEEEproof}

Proposition \ref{pro4} shows that for this case, the computation rate affects the migration decision through the minimum channel threshold $h_n^{\rm{min}}$ [see \eqref{eqn:hnmin}], with $h_n^{\rm{min}}$ being decreasing as $f_n(k)$ increases, and the migration policy is similar to that of the preceding case but works on the basis of $\nu_n(k)$.

\vspace{-0.1in}
\subsection{Performance Analysis}
In this subsection, we present the performance bounds of the proposed algorithm. For ease of analysis, we assume that the system randomness is i.i.d. over frames and that \makebox{Problem (P1)} is feasible. As such, the feasibility implies that there exists a slack constant $\delta>0$ and a feasible solution to \makebox{Problem (P1)} such that the following inequality holds for all $k$:
\begin{align}\label{eqn:thm2_condi}
\frac{1}{T} \,\mathbb{E} \Big\{\sum_{t\in\mathcal{T}_k}X(t)\Big\} < \epsilon - \delta.
\end{align}

Based on this, we have the following theorem for theoretically quantifying the performance bounds that the proposed algorithm can achieve.
\begin{theorem}\label{pro5}
Assume that the condition \eqref{eqn:thm2_condi} is satisfied for $\exists \delta>0$, and the initial virtual queue length is zero, i.e., $Q(0)=0$. Then, for any $V>0$, we have:
\begin{enumerate}[1)]
\item The average queue length under the proposed algorithm is upper bounded by
\begin{align}\label{eqn:thm21}
\lim_{K\rightarrow\infty} \frac{1}{K} \sum^{K-1}_{k=0} \mathbb{E} \{Q^*(kT)\} \leq \frac{B_2 +VE_{\rm{max}}}{\delta},
\end{align}
where $Q^*(t)$ denotes the resultant queue length by the proposed algorithm and $E_{\rm{max}}\triangleq \overline{P}\cdot\max_{n\in\mathcal{N}}\big\{\tau_d-\tfrac{\xi }{f_{n}^\text{max}}\big\}$.
\item The average energy consumption achieved by the proposed algorithm satisfies
\begin{align}\label{eqn:thm22}
\lim_{K\rightarrow \infty} \frac{1}{KT}\sum_{k=0}^{K-1}\sum_{t\in\mathcal{T}_k} \mathbb{E}\left\{\mathcal{E}^{\text{*}}(t)\right\}\leq \mathit{E}_{\rm{av}}^{\emph{opt}} + \frac{B_2}{V},
\end{align}
where $\mathcal{E}^{\text{*}}(t)$ denotes the resultant energy consumption by the proposed algorithm and $\mathit{E}_{\rm{av}}^{\emph{opt}}$ denotes the minimum average energy consumption for Problem (P1).
\end{enumerate}
\end{theorem}
\begin{IEEEproof}
See Appendix F.
\end{IEEEproof}

Theorem \ref{pro5} shows that the average energy consumption of the online algorithm can asymptotically achieve the optimum $\mathit{E}_{\rm{av}}^{\emph{opt}}$  of the original Problem (P1) by increasing the control parameter $V$. Besides, the average virtual queue length is bounded by $\mathcal{O}(V)$ in \eqref{eqn:thm21}, indicating the queue is mean rate stable and the reliability constraint \eqref{eqn:st1} is guaranteed.

\section{Extension to Multiuser Management}
In this section, we consider the multiuser mobility management under the proposed two-timescale framework. Multiuser migrations could noticeably change the load of BSs and affect the computation rates for other users associated at the same BS. Thus, compared with the single-user case, multiuser management requires considering the load balance factor among BSs when making users' migration/association decisions.

Specifically, we consider that $M$ users, denoted by set $\mathcal{M} = \{1,\cdots, M\}$, are randomly distributed and move in the multi-cell network. We assume that each user is allocated with a dedicated channel (like in OFDMA) for multiuser offloading \cite{Lyapunov3,predict1,multiuser1}. Regarding the multiuser computing, we use the number of associated users $y_n(k)$ to represent the load of BS $n$ in the $k$-th frame, and model the computation rate for user $i$ as a function of $y_n(k)$ \cite{parallel_computing_model, ourwork2}, which are respectively given by\footnote{Other load-aware computation models, such as equal resource allocation among the users at a BS, are also applicable to our proposed management scheme. }
\begin{align}\label{eqn:fnik}
\begin{cases}
y_n(k) = \sum\limits_{i\in\mathcal{M}}\mathds{1}_{\{n_i(k) = n\}}, &\forall n\in\mathcal{N},\\
f_{i,n}(k) = F_{i,n}\  \alpha_n^{\ y_n(k) -1}, &\forall n\in\mathcal{N},  \quad \forall i\in\mathcal{M},
\end{cases}
\end{align}
where $n_i(k)$ denotes the association decision of user $i$ and $\alpha_n \in (0,1)$ is the degradation factor that specifies the percentage decrease of user's computation rate as the BS load increases.

The models of task arrival, service migration, and two-timescale operation follow the same settings in the single-user case for each user. Our goal in multiuser management is to minimize the sum of users' time-averaged energy consumption while guaranteeing the reliability requirement of each user, which is formulated as
\begin{subequations}
\begin{align}
(\text{P2})~~\min_{\{n_i(k)\},  \{p_i(t)\}} \quad &\lim_{K\rightarrow \infty} \frac{1}{KT}\sum_{k=0}^{K-1}\sum_{t\in\mathcal{T}_k}\sum_{i\in\mathcal{M}} \mathbb{E}\left\{\mathcal{E}_i(t)\right\} \label{eqn:ob2} \\
\rm{s.t.} ~~\quad ~~&\lim_{K\rightarrow \infty} \frac{1}{KT}\sum_{k=0}^{K-1}\sum_{t\in\mathcal{T}_k} \mathbb{E}\left\{X_i(t)\right\}\leq \epsilon_i, \quad \forall i\in \mathcal{M},\label{eqn:P2st1}\\
&~n_i(k)\in \mathcal{N}, \quad \quad \ \ \forall i\in \mathcal{M},\quad k = 0, 1,  \cdots \label{eqn:P2st2}\\
&~0\leq p_i(t)\leq \overline{P}_i, \quad\forall i\in \mathcal{M}, \quad  t = 0, 1,  \cdots \label{eqn:P2st3}\\
&\eqref{eqn:fnik} \text{\ for per-frame computation resource allocation.} \quad \quad \nonumber
\end{align}
\end{subequations}
Apart from the addition of subscript $i$ to denote the user index, all notations in the above Problem (P2) and their corresponding expressions remain the same as the single-user case.

Similarly, we can develop the Lyapunov-based online algorithm to solve Problem (P2). The algorithm framework is similar to Algorithm \ref{alg:1} (see Section IV-B) and consists of three control actions: multiuser migration decisions per frame, user's power control per slot, and the virtual queue update. The last two actions are executed on each user in parallel and consistent with the results of the single-user case, i.e., each user carries out the power strategy in Proposition \ref{pro1} and update its queue according to \eqref{eqn:virtualqueue} at every slot. Thus, in what follows we focus on the per-frame multiuser migration problem.

\vspace{-0.1in}
\subsection{Per-frame Multiuser Migration Problem}
At the beginning of each frame $k$, based on the observation of $\{Q_i(kT)\}$ and $\{n_i(k-1)\}$, the network operator decides the users' associations $\{n_i(k)\}$ by solving the following per-frame problem:
\begin{align}\label{eqn:perframeproblem2}
\min_{\{n_i(k)\}, \{p_i(t)\}} \quad&\mathbb{E}\bigg\{\sum_{t\in \mathcal{T}_k}\sum_{i\in\mathcal{M}}V\mathcal{E}_i(t)+Q_i(kT)X_i(t)\bigg\}  \\
\rm{s.t.} ~~\quad ~~&\eqref{eqn:fnik}, \eqref{eqn:P2st2}, \eqref{eqn:P2st3}. \nonumber
\end{align}
Note that the user's computation rate in Problem \eqref{eqn:perframeproblem2} is no longer a known constant but a function of $\{n_i(k)\}$  due to the constraints \eqref{eqn:fnik}. Thus the users' association decisions $\{n_i(k)\}$ are coupled in the per-frame Problem \eqref{eqn:perframeproblem2}.

To facilitate exposition, we introduce a set of binary variables $\mathbf{X} = \{x_{i,n}\}$ to represent the users' association decisions $n_{i}(k)$, with $x_{i,n} = 1$ indicating $n_{i}(k) = n$ and $x_{i,n} = 0$ otherwise. Then, by incorporating the power strategy in Proposition \ref{pro1} and taking the channel assumptions as in the single-user case, i.e., i.i.d. channel randomness over the slots of a frame and the channel statistics in a frame being known, we can refine Problem \eqref{eqn:perframeproblem2} as the following problem:
\begin{subequations}\label{eqn:perframeproblem3}
\begin{align}
\min_{\mathbf{X}\in\{0, 1\}^{M\times N} } ~~&\ R(\mathbf{X})= \sum_{i\in\mathcal{M}}\sum_{n\in\mathcal{N}} x_{i,n} Z_{i,n}^{\emph{sum}}(y_n) \label{eqn:perframeproblem31} \\
\rm{s.t.} \quad~~&\sum_{i\in\mathcal{M}} x_{i,n} = y_n, \quad \forall n \in \mathcal{N}, \label{eqn:perframeproblem32}\\
&\sum_{n \in \mathcal{N}} x_{i,n} = 1,  ~\ \quad \forall i \in \mathcal{M}. \label{eqn:perframeproblem33}
\end{align}
\end{subequations}
Here we drop the frame index $k$ here for ease of notation. $Z_{i,n}^{\emph{sum}}(y_n)$ in \eqref{eqn:perframeproblem31} is the minimum expected cost of user $i$ if associating with BS $n$ in the current frame given the BS load $y_n$ (that determines $f_{i,n}$). The expression of $Z_{i,n}^{\emph{sum}}(y_n)$ is the same as $Z_{n}^{\emph{sum}}(k)$ of the single-user case and given by \eqref{eqn:Znk_sum}. $R(\mathbf{X})$ is the sum of users' association cost, which can be regarded as the overall system cost. The constraints \eqref{eqn:perframeproblem32} and \eqref{eqn:perframeproblem33} are equivalent to \eqref{eqn:fnik} and \eqref{eqn:P2st2}, respectively.

Note that due to the binary variables $\mathbf{X}$, Problem \eqref{eqn:perframeproblem3} is an integer nonlinear programming problem that is hard to obtain the optimal solution in efficient time-complexity. For this reason, we find a near-optimal solution by developing a low-complexity algorithm in the next subsection.

\vspace{-0.1in}
\subsection{Algorithm Design for Multiuser Migration}
We propose an efficient iterative algorithm which converges to a near-optimal solution to the migration Problem \eqref{eqn:perframeproblem3}. The algorithm is based on the intuition that, the user with the worse BS association is more likely to trigger migration to another BS (with a stronger wireless link and/or less compute load), which consequently reduces association cost of the user and the system. Motivated by this, we design an algorithm centering on the worst user-BS association improvement.

\begin{algorithm}[t]
\caption{Suboptimal Algorithm for Solving Problem \eqref{eqn:perframeproblem3}}
\begin{algorithmic}[1]\label{alg:2}
\STATE Initialize $l = 1$, $s = 1$, and set $\mathbf{X}^{(1)}$ as the current user-BS associations.
\STATE Find the worst user-BS association $(i^{(1)}, n^{(1)})$ by \eqref{eqn:worstpair}.
\REPEAT
\STATE Generate $(N-1)$ new association matrixes $\mathbf{X}_{m}^{(l)}$, $\forall m\in \mathcal{N}\backslash n^{(l)}$, each by switching the $i^{(l)}$-th user's association from BS $n^{(l)}$ to BS $m$.
\STATE Update $\mathbf{X}^{(l+1)}$ by \eqref{eqn:Xupdate}.
\STATE Find the worst user-BS association $(i^{(l+1)}, n^{(l+1)})$ by \eqref{eqn:worstpair}.
    \IF{$\mathbf{X}^{(l+1)} = \mathbf{X}^{(l)}$}
        \STATE Let $i^\prime \leftarrow \text{mod}(s-1, M)+1$ and $n^\prime$ be the index of the $i^\prime$-th user associated BS.
        \STATE $(i^{(l+1)}, n^{(l+1)})\leftarrow (i^\prime, n^\prime)$.
        \STATE $s\leftarrow s+ 1$.
    \ENDIF
\STATE $l \leftarrow l +1$.
\UNTIL $\mathbf{X}^{(l)}$ remains unchanged after $M$ consecutive iterations.
\ENSURE $\mathbf{X}^{(l)}$.
\end{algorithmic}
\end{algorithm}

The algorithm for solving Problem \eqref{eqn:perframeproblem3} is presented in Algorithm \ref{alg:2}. It starts by initializing the user-BS associations $\mathbf{X}^{(1)}$ and finding the worst user-BS association $(i^{(1)}, n^{(1)})$ from $\mathbf{X}^{(1)}$ as described later via \eqref{eqn:worstpair}. At each iteration $l$, the algorithm goes through the following two steps:

\subsubsection{Association update} In this step, we adjust the association decision of user $i^{(l)}$, which is equal to update the whole association matrix $\mathbf{X}^{(l)}$ under the entries of other users being fixed. Specifically, we generate $(N-1)$ new association matrixes $\mathbf{X}^{(l)}_m$, each representing user $i^{(l)}$ is migrated to other BS $m$, with $m\in \mathcal{N}\backslash n^{(l)}$. Then, among the current and new association matrixes, we choose the one with the minimum system cost $R(\mathbf{X})$ [see \eqref{eqn:perframeproblem31}] as the best association matrix for the next iteration:
\begin{align}\label{eqn:Xupdate}
\mathbf{X}^{(l+1)} = \arg \min\limits_{\mathbf{X}^{(l)}, \{\mathbf{X}^{(l)}_m|\forall m\in \mathcal{N}\backslash n^{(l)}\}} R(\mathbf{X}).
\end{align}
We can observe from \eqref{eqn:Xupdate} that if $\mathbf{X}^{(l+1)}\neq \mathbf{X}^{(l)}$, the system cost $R(\mathbf{X})$ [i.e., the objective value of Problem \eqref{eqn:perframeproblem3}] is always decreasing in the association update.

Next, based on the updated $\mathbf{X}^{(l+1)}$, we find the worst user-BS association $(i^{(l)}, n^{(l)})$ for the next iteration by comparing the users' association costs:
\begin{align}\label{eqn:worstpair}
(i^{(l+1)}, n^{(l+1)}) = \arg\max\limits_{i,n} \left\{x_{i,n}^{(l+1)} Z_{i,n}^{\emph{sum}}(y_n^{(l+1)}) \right\},
\end{align}
where $y_n^{(l+1)}$ is computed according to \eqref{eqn:perframeproblem32}.

\subsubsection{User switching} When $\mathbf{X}^{(l+1)}=\mathbf{X}^{(l)}$, it means that the system cost can not be further reduced by adjusting the association of the worst user $i^{(l)}$. Furthermore, the worst users are equal (i.e., $i^{(l+1)}= i^{(l)}$) in the following iterations, leading to no more changes in the system cost. Clearly, in order to find a potential system-cost reduction, we need to switch another user to adjust its association. To this end, we introduce a user switching step. At each switching step $s$, let $i^\prime =\text{mod}(s-1, M)+1$, and we select user $i^\prime$ instead of the current worst user $i^{(l+1)}$ to perform the association update in the next iteration. Note that the switching process does not affect the non-increasing property of the system cost in the association update.

The iteration process is repeated until $\mathbf{X}^{(l)}$ remains unchanged after $M$ consecutive iterations. The convergence is guaranteed because the system cost always keeps non-increasing in iterations, and all $M$ users have been swept by the switching process and have no association changes when the stopping condition is met.

\section{Simulation Results}
The simulation settings are as follows. We consider that $N=25$ BSs are regularly deployed in a $2$ km$\times$2 km square area. The slot length and the frame size are set to be $\tau=10$ ms and $T=500$ slots, respectively. The time horizon is $K=2500$ frames. We consider a task type with $L=5$ Kbits, $\xi=2640$ cycles/bit (such as $400$ frame video game \cite{multiuser1}), $\tau_d=10$ ms, and $\rho=0.5$. In terms of the user movement, we assume that the user's locations change over frames and adopt the Random Waypoint Mobility model \cite{RWP} to generate the user's location for each frame, with the parameters taken as: the static probability and pause time $p_s=t_p=0$, and the user's velocity $v= v_\text{min}=v_\text{max}=5$ m/s. For task offloading, the channel power gains are modeled as $h_n(t)= g_n(t)H_n(k)$ in \eqref{eqn:hnt}. The large-scale fading $H_n(k)$ is given by $127+30\log_{10}(10^{-3} d_n(k))$, where $d_n(k)$ denotes the distance between the user and BS $n$ in meter at the $k$-th frame. The small-scale fading  $g_n(t)$ follows normalized exponential distribution. Besides, the noise power spectrum density is set as $-174$ dBm with $W=10$ MHz channel bandwidth. For service migration, we consider $C=5$, i.e., migration/handover delay is $50$ ms \cite{handoverdelay}. Unless mentioned otherwise, the main communication and computation parameters used in the simulations are summarized in Table \ref{tab:1}.

\begin{table}[t]
\setlength{\abovecaptionskip}{0.cm}
\setlength{\belowcaptionskip}{-0.4cm}
\centering
\caption{\label{tab:1}System Parameters}
\begin{tabular}{|c|c|}
\hline
Parameter & Value \\
\hline
Number of BSs, $N$   & $25$   \\
\hline
Slot length, $\tau$  & $10$ ms \\
\hline
Frame size, $T$ & $500$ slots \\
\hline
Task arrival probability, $\rho$ & $0.5$ \\
\hline
Peak transmit power, $\overline{P}$   &  $1$ W \\
\hline
~BS computation rate, $f_n(k)$ ~& $[1\times10^{10}, 2\times10^{10}]$ cycles/s\\
\hline
Service migration delay, $C$ & $5$ slots  \cite{handoverdelay} \\
\hline
Reliability threshold, $\epsilon$ & $1\times 10^{-3}$  \\
\hline
Control parameter, $V$ & $5000$\\
\hline
\end{tabular}
\end{table}

For performance comparison, we also simulate the two traditional handover schemes as the benchmarks:
\begin{enumerate}
\item \emph{Received signal strength (RSS) only:} For each frame, the user always migrates the association to the BS with the highest large-scale fading $H_n(k)$.
\item \emph{RSS plus hysteresis:} Let $n^\prime\triangleq \arg\max_{n\in\mathcal{N}\backslash n(k-1)}\left\{H_n(k)\right\}$ denote the target BS with the highest $H_n(k)$ at $k$-th frame. For each frame, the user migrates to the target BS if $H_{n^\prime} (k) >  (1+ \beta_\text{th}) H_{n(k-1)}(k)$; otherwise it stays at the current BS. Here, $\beta_\text{th}$ is a hysteresis margin and set as $2$ in the simulations.
\end{enumerate}

Note that the two benchmarks are used to decide the service migration for each frame. For per-slot offloading of the arrived tasks, we consider that both of them use the following power strategy:
\begin{align}\label{eqn:p_bench}
p(t)=\begin{cases}
p_{n(k)}^{\rm{min}}(t), & \text{if } t\in \mathcal{T}_k\backslash \mathcal{T}_k^c \text{~and~} p_{n(k)}^{\rm{min}}(t) \leq p^{\rm{max}}(t), \\
0, &\text{otherwise},
\end{cases}
\end{align}
where $p_{n(k)}^{\rm{min}}(t)$ is the minimum required power as in \eqref{eqn:pn_min} and $p^{\rm{max}}(t)\triangleq \overline{P}\cdot\mathds{1}_{\left\{\mathit{X}_{\rm{av}}(t)> \epsilon\right\}} + 0.05\overline{P}\cdot\mathds{1}_{\left\{\mathit{X}_{\rm{av}}(t)\leq \epsilon\right\}}$ is the maximum power threshold at slot $t$, which is set to be $\overline{P}$ or $0.05\overline{P}$ depending on whether the current average task-failure rate $\mathit{X}_{\rm{av}}(t)$ exceeds the \makebox{threshold $\epsilon$} or not. Here, an online reliability control is made in \eqref{eqn:p_bench} by setting two modes on $p^{\rm{max}}(t)$:  $0.05\overline{P}$ that prefers energy saving, and $\overline{P}$ that emphasizes more on reliability.

\begin{figure}[t]
\setlength{\belowcaptionskip}{-0.4cm}
\centering
\subfigcapskip=-10pt
\subfigure[]{\label{sub.1}
    \includegraphics[width=0.47\textwidth]{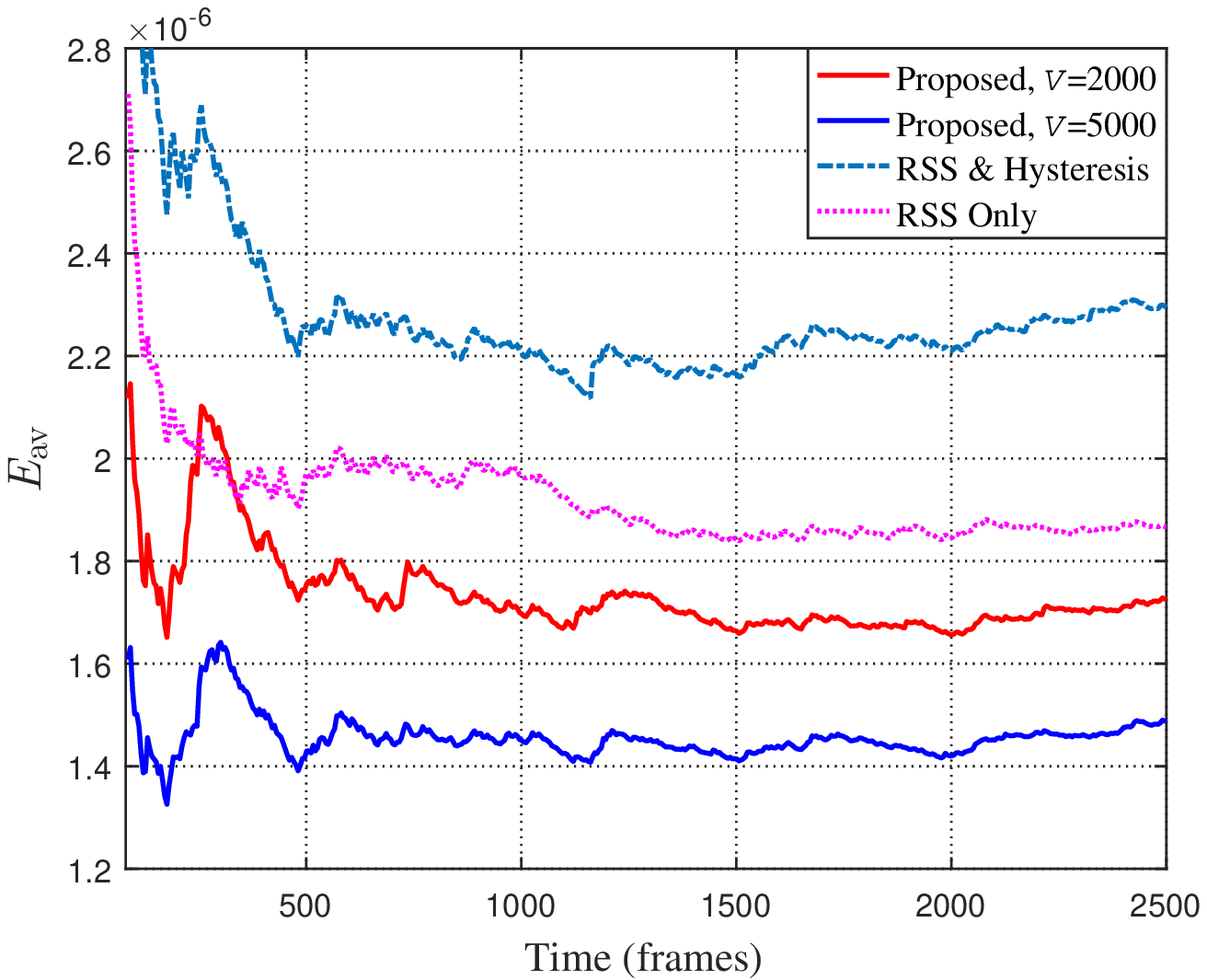}
}
\subfigure[]{\label{sub.2}
    \includegraphics[width=0.47\textwidth]{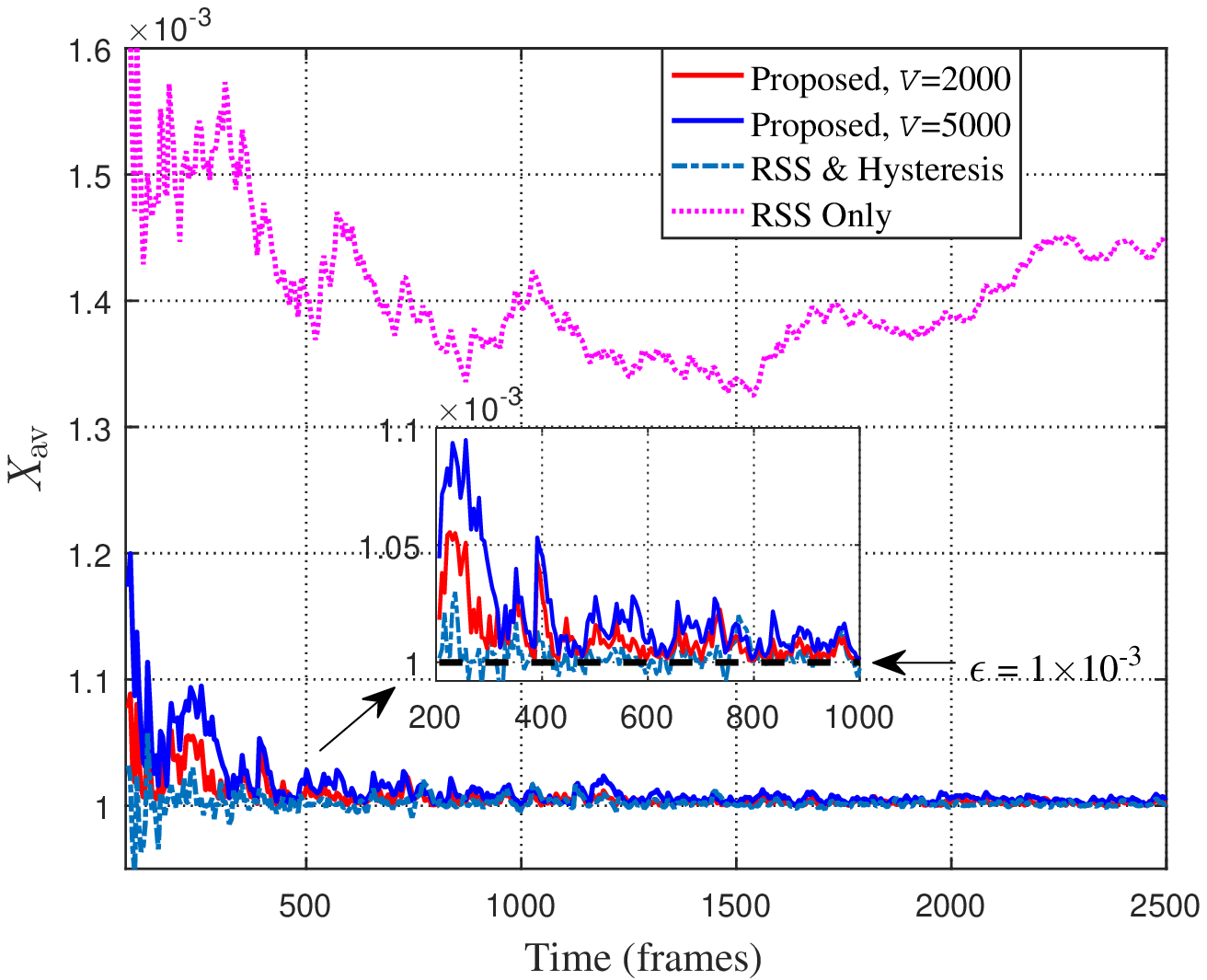}
}
\vspace{-0.1in}
\caption{Time evolution: (a) average energy consumption, and (b) average task-failure rate.}
\label{fig:3}
\end{figure}

\vspace{-0.1in}
\subsection{Single-user Case}
Fig. \ref{fig:3} shows the average energy consumption and task-failure rate of the proposed \mbox{Algorithm \ref{alg:1}} and two benchmark schemes over $2500$ time frames. First, it can be seen that our proposed algorithm in $V=2000$ and $V=5000$ both can achieve lower energy consumption than the two benchmarks while satisfying the reliability constraint. A larger value $V$ in the proposed algorithm can save more energy; however, as shown in the local diagram of Fig. \ref{fig:3}(b), its task-failure rate converges more slowly to the reliability threshold $\epsilon$. Among the benchmarks, we can observe that the RSS only scheme has lower energy consumption but does not meet the reliability constraint caused by frequent service migrations; in contrast, the RSS plus hysteresis can reduce excessive migrations to ensure the reliability but at the expense of high energy consumption due to its delayed migration response.

\begin{figure}
\setlength{\belowcaptionskip}{-0.5cm}
\centering
  \begin{minipage}[t]{0.5\linewidth}
    \centering
    \includegraphics[width=\textwidth]{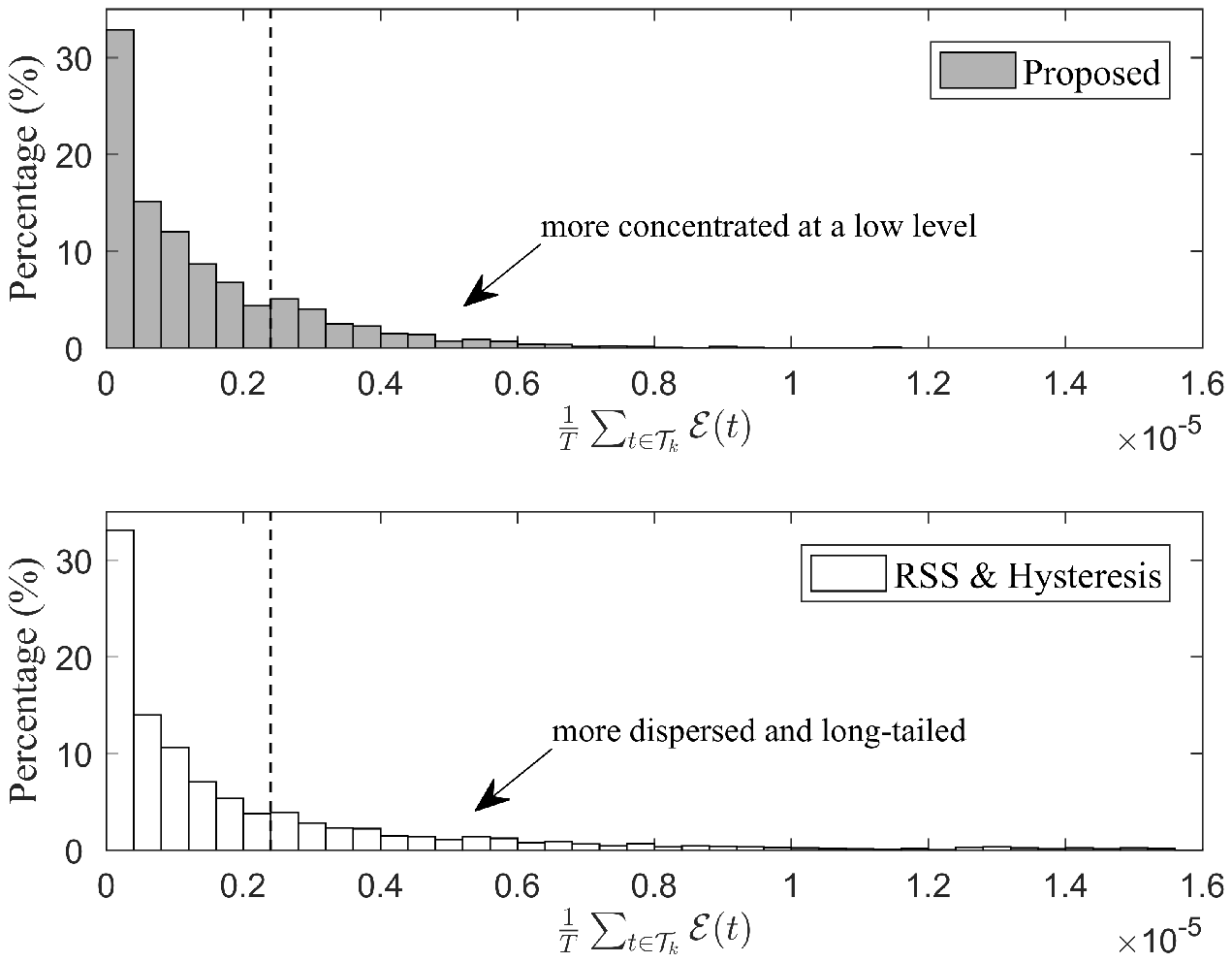}
    \caption{Distribution of per-frame average energy consumption.}
    \label{fig:distribution}
  \end{minipage}%
  \hfill
  \begin{minipage}[t]{0.5\linewidth}
    \centering
    \includegraphics[width=\textwidth]{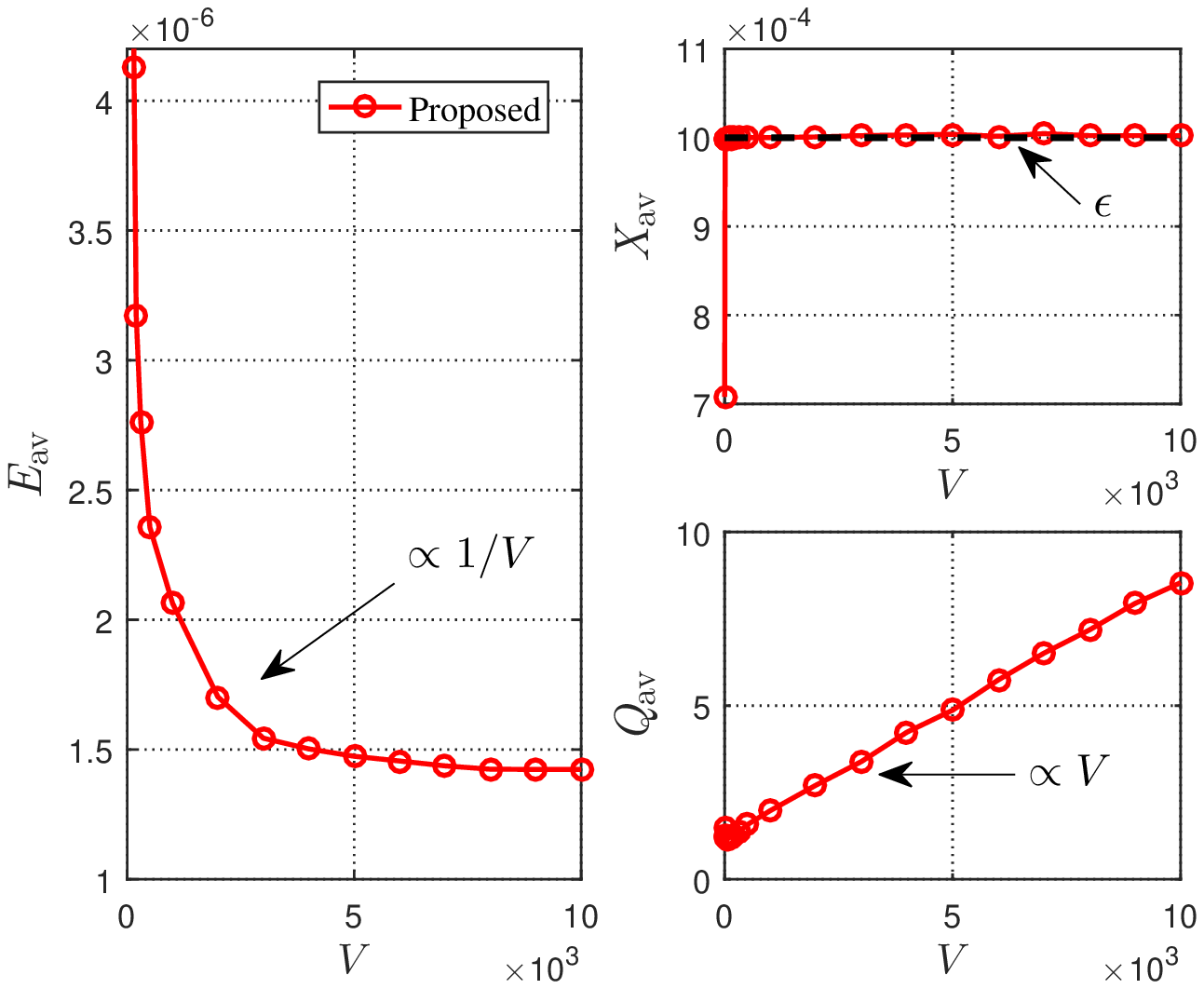}
    \caption{Impact of control parameter $V$.}
    \label{fig:V}
  \end{minipage}
\end{figure}

In Fig. \ref{fig:distribution}, we compare the distribution of per-frame average energy consumption between the RSS plus hysteresis scheme and the proposed Algorithm \ref{alg:1} with $V=5000$. It can be observed that, during the interval $[0.22\times 10^{-5}, \infty)$, the energy distribution of the proposed algorithm is more centralized at a low level, while the distribution of the RSS plus hysteresis is dispersed and long-tailed at high energy level (e.g., $[0.6\times 10^{-5}, \infty)$). Note that the service migration mainly serves for energy reduction at the high energy interval corresponding to the user's locations at the cell edge. Therefore, this demonstrates that compared to the RSS plus hysteresis scheme, our proposed algorithm can make more accurate and prompt migration decisions when the user moves across the BSs to reduce energy consumption.

Fig. \ref{fig:V} shows the impact of control parameter $V$ on the average energy consumption, the task-failure rate, and the virtual queue length of our proposed Algorithm \ref{alg:1}, where $\epsilon=10^{-3}$. We can see that the energy consumption decreases inversely proportional to $V$; the task-failure rate maintains satisfying the reliability constraint no matter what $V$ is; and the average queue length increases linearly as $V$ increases. These match the results in Theorem \ref{pro5} that the performance of average energy consumption and queue length follows the $[\mathcal{O}(1/V), \mathcal{O}(V)]$ tradeoff.

\begin{figure}
\centering
  \begin{minipage}[t]{0.5\linewidth}
    \centering
    \includegraphics[width=\textwidth]{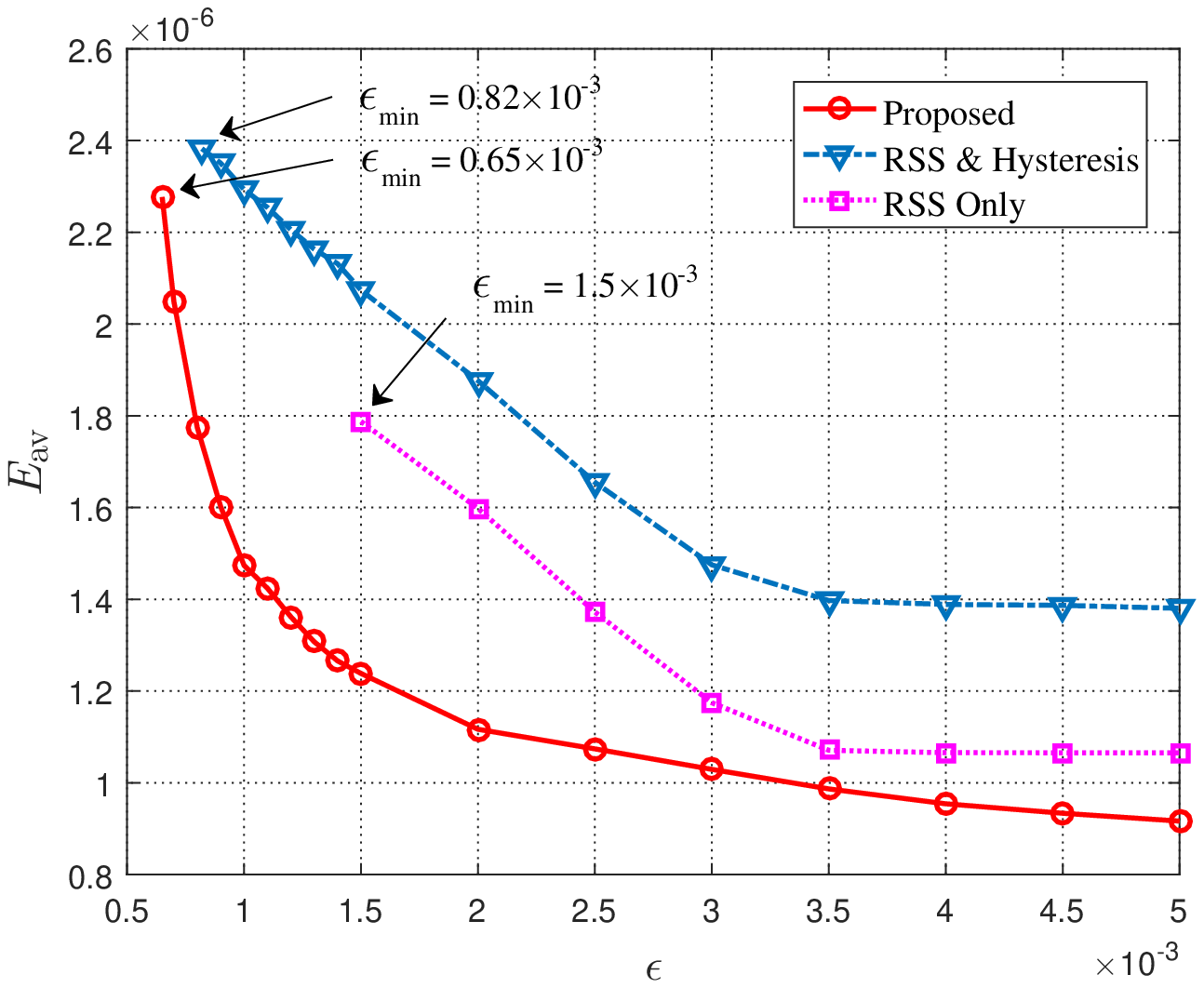}
    \caption{Average energy consumption vs. $\epsilon$.}
    \label{fig:tradeoff}
  \end{minipage}%
  \hfill
  \begin{minipage}[t]{0.5\linewidth}
    \centering
    \includegraphics[width=\textwidth]{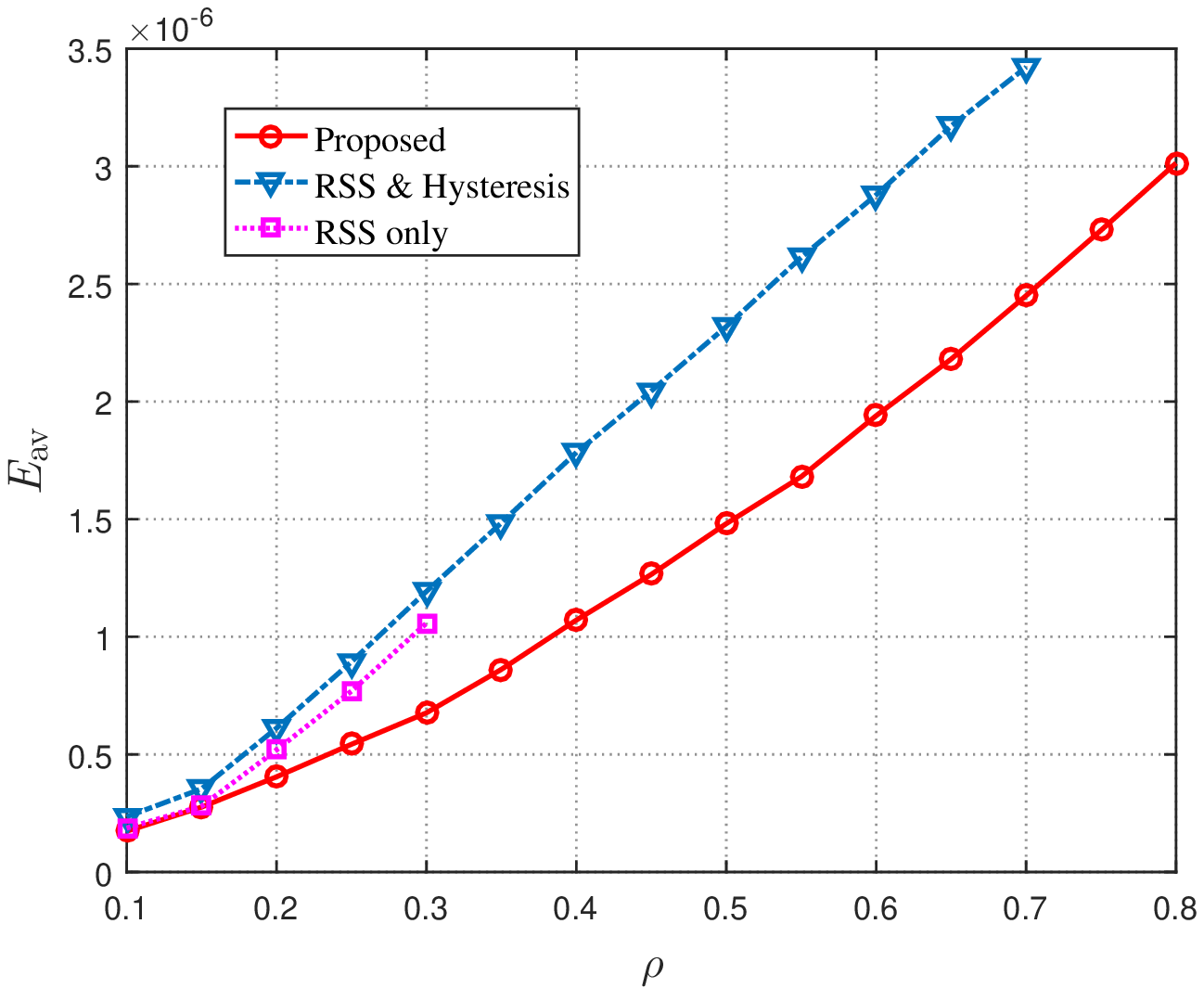}
    \caption{Average energy consumption vs. $\rho$.}
    \label{fig:rho}
  \end{minipage}
\end{figure}

Fig. \ref{fig:tradeoff} shows the energy-reliability tradeoff of all the algorithms by varying the \makebox{threshold $\epsilon$}. We can observe that the proposed Algorithm \ref{alg:1} always achieves a smaller energy consumption than the two benchmarks under the same reliability requirement. The RSS only scheme performs well when $\epsilon>3\times 10^{-3}$. This is because when $\epsilon$ is large, the reliability loss in migration is tolerable and migrating the BS with the best channel for each frame helps reduce user's (transmit) energy consumption. However, it fails to fulfill the stringent reliability requirement due to its aggressive migration strategy. The RSS plus hysteresis can enhance the reliability performance compared to the RSS only, but it suffers high energy consumption. In contrast, our proposed algorithm outperforms the two benchmarks in both reliability and energy performance due to its joint management of service migration and computation offloading.

Fig. \ref{fig:rho} shows the average energy consumption versus the task arrival probability $\rho$. As expected, the proposed Algorithm \ref{alg:1} achieves significant energy reduction compared to the two benchmarks under the same $\rho$. In addition, our proposed algorithm can accommodate a higher task arrival rate $\rho$ than the two benchmarks to meet the reliability constraint.

\begin{figure}
\centering
  \begin{minipage}[t]{0.5\linewidth}
    \centering
    \includegraphics[width=\textwidth]{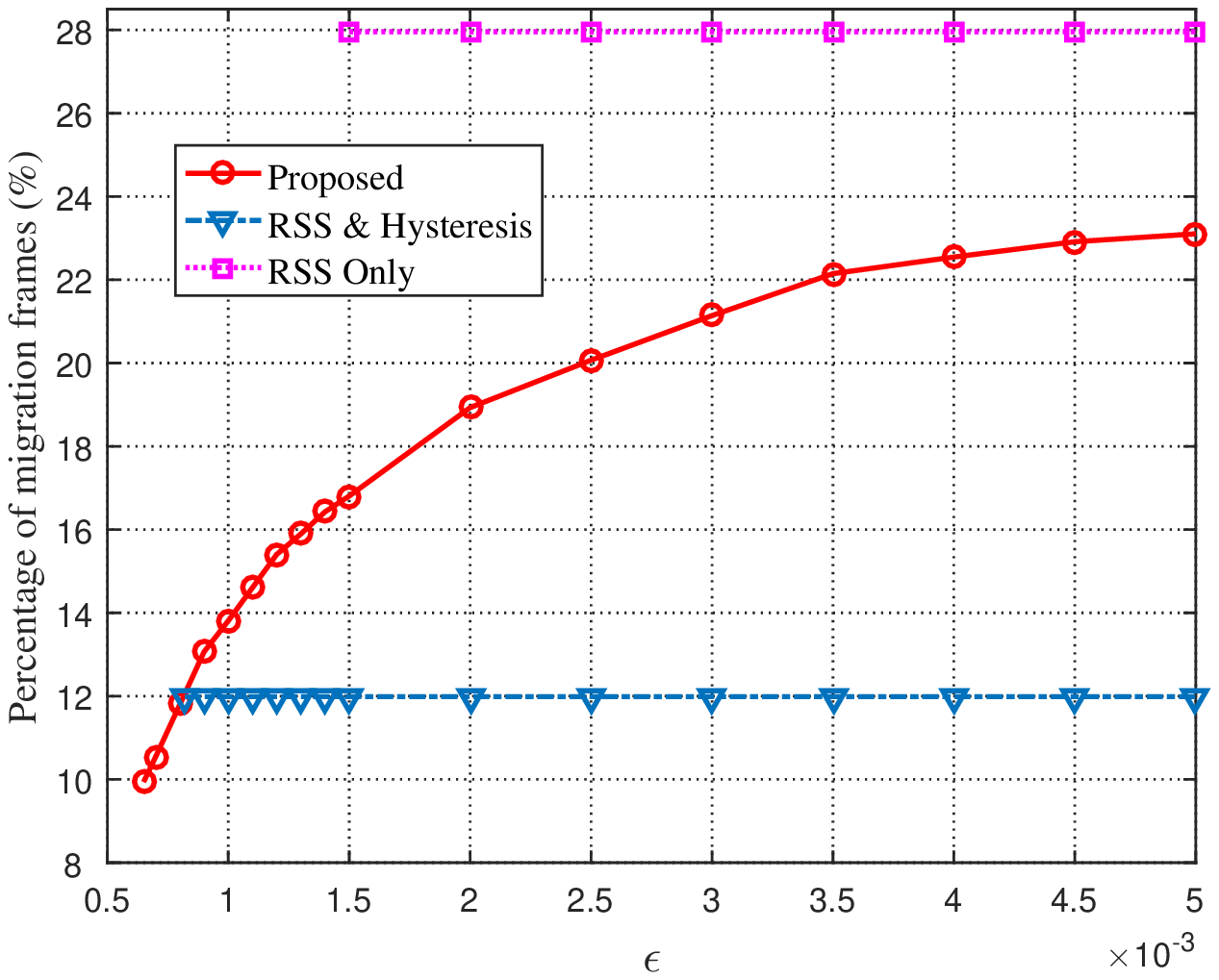}
    \caption{Percentage of migration frames vs. $\epsilon$.}
    \label{fig:migpercent}
  \end{minipage}%
  \hfill
  \begin{minipage}[t]{0.5\linewidth}
    \centering
    \includegraphics[width=\textwidth]{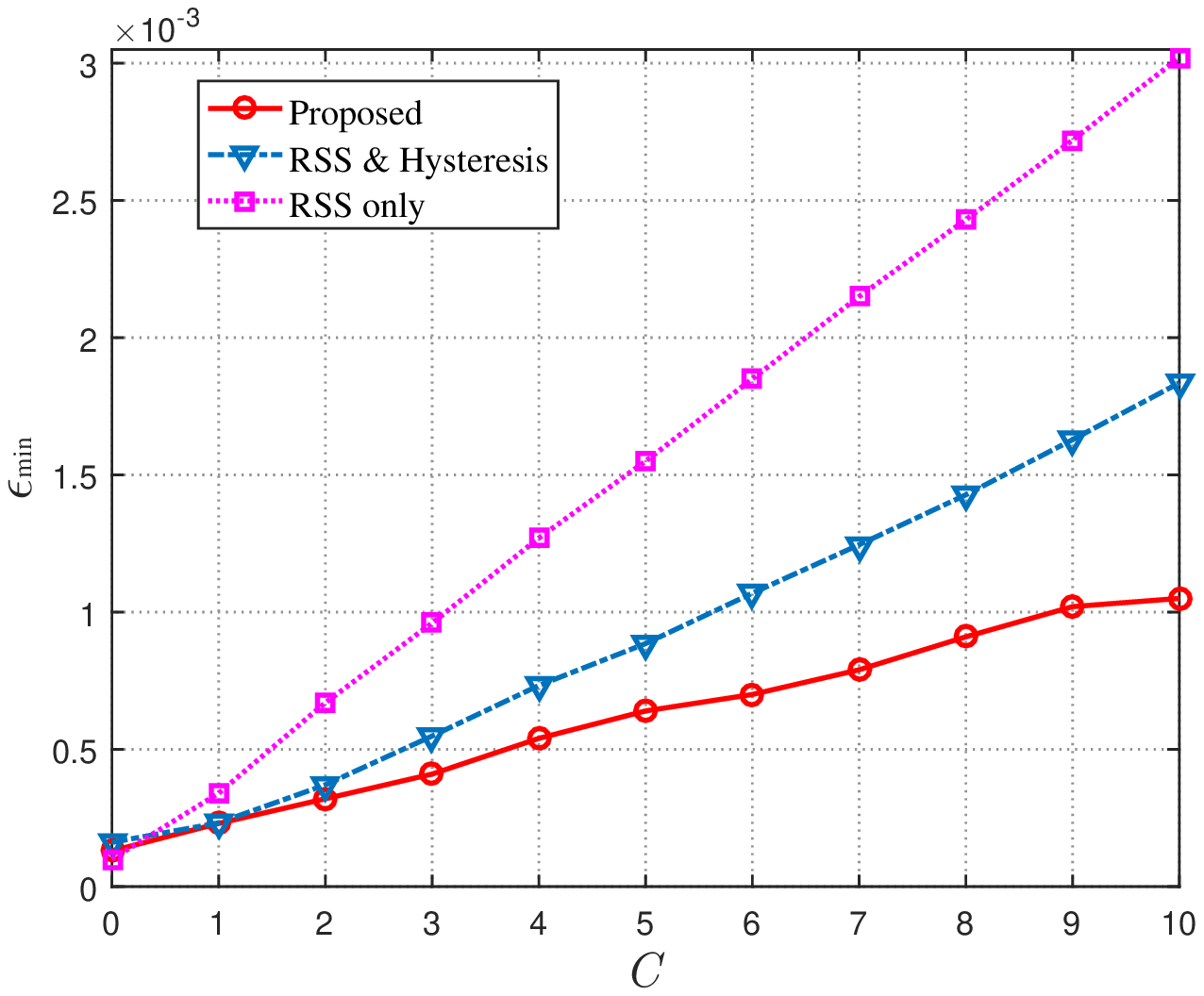}
    \caption{Minimum reliability threshold $\epsilon_\text{min}$ vs. $C$. }
    \label{fig:cmax}
  \end{minipage}
\end{figure}

In Fig. \ref{fig:migpercent}, we plot the percentage of migration frames among the total $K=2500$ frames under different reliability thresholds $\epsilon$. As we observe, the RSS only and the RSS plus hysteresis have static migration percentage since their migration policies are only related to channel condition, while the proposed Algorithm \ref{alg:1} can adjust the migration percentage according to the reliability requirement. Combining with the energy behaviors as shown in Fig. \ref{fig:tradeoff}, these demonstrate that our proposed algorithm performs more flexible migration-frequency control to balance the energy consumption and the reliability performance.

Finally, we evaluate the impact of service migration delay $C$ on the reliability performance in Fig. \ref{fig:cmax}, where the reliability performance is measured by the minimum threshold $\epsilon_\text{min}$ that can be supported by the algorithms. First, we can see that $\epsilon_\text{min}$ is increasing with $C$ in all considered algorithms while the ascending rate of our proposed Algorithm \ref{alg:1} is the slowest, indicating that the proposed has the best reliability performance against the migration-delay effect. We also observe that the performance of two benchmarks is  close to that of the proposed Algorithm \ref{alg:1} when $C$ is small; however, they dramatically deteriorate as $C$ becomes large. This is because the reliability loss (i.e., task failure) caused by migration is low when $C$ is small, while it becomes a dominant factor and requires effective management when $C$ goes large.

\vspace{-0.1in}
\subsection{Multiuser Case}
In this subsection, we verify the performance of our proposed Algorithm \ref{alg:2} in multiuser management. Similarly, we use the Random Waypoint Mobility model to generate the movement of multiple users, where each user moves in a constant speed $v_i$ (m/s), which is randomly chosen from the set $\{2, 4, 6, 8, 10\}$. For multiuser computing, we set $F_{i,n} = 2 \times 10^{10}$ cycles/s and the degradation factor $\alpha_n = 0.926$. Other parameters for each user follow the same settings of the single-user case.

\begin{figure}
\centering
  \begin{minipage}[t]{0.5\linewidth}
    \centering
    \includegraphics[width=\textwidth]{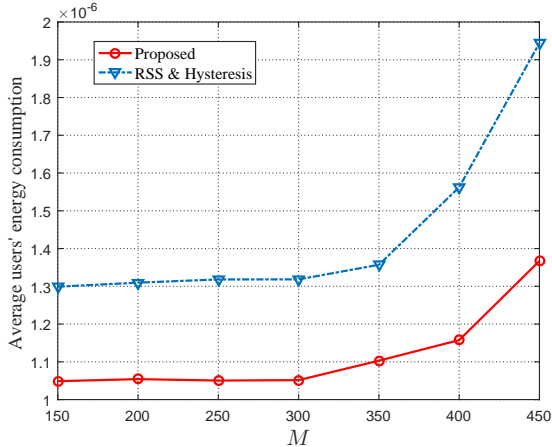}
    \caption{Average users' energy consumption vs. $M$.}
    \label{fig:multiuserM}
  \end{minipage}%
  \hfill
  \begin{minipage}[t]{0.5\linewidth}
    \centering
    \includegraphics[width=\textwidth]{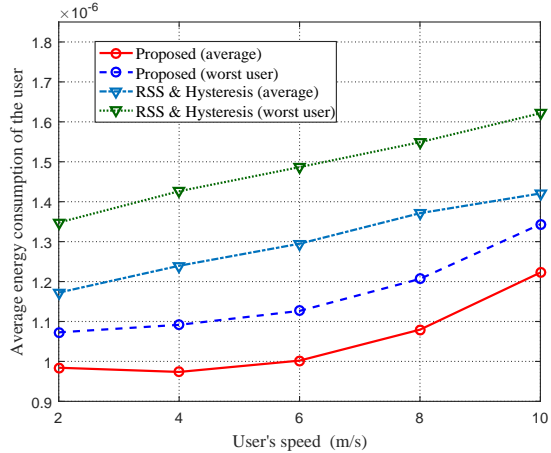}
    \caption{Average energy consumption of the users with different speeds in both the worst and average cases. }
    \label{fig:multiuserV}
  \end{minipage}
\end{figure}

In Fig. \ref{fig:multiuserM}, we plot the average users' energy consumption versus the number of users in the network $M$, under the proposed Algorithm \ref{alg:2} and the benchmark scheme of RSS plus hysteresis. We can observe that for both the proposed and the benchmark, the average users' energy consumption is insensitive to the increase of $M$ when $M < 300$ while it begins to increase when $M > 300$. The reason is that, when $M$ is small, each BS is lightly loaded and can provide stable computation rates; when $M$ becomes large, the computation rate suffer severe degradation due to the overloaded BS. Nevertheless, the energy consumption of our proposed algorithm increases at slower rate than that of the benchmark scheme, thanks to its efficient load-aware migration mechanism to balance the workload among BSs.

Fig. \ref{fig:multiuserM} shows the influence of user's speed on its average energy consumption, where the average performance and the performance of the worst user are considered, and $M$ is set as $250$ for this case. We can see that the energy consumption increases with the user's speed in both the propose Algorithm \ref{alg:2} and the RSS plus hysteresis scheme, due to the growth of migration demands. Compared to the RSS plus hysteresis scheme, our proposed algorithm has lower energy consumption and a smaller gap between the average and the worst-user performance. The first one is because our proposed algorithm can make more accurate and prompt migration decisions for every user as discussed in Fig. \ref{fig:multiuserM}. The second one is because our multiuser migration strategy in the proposed algorithm centres on  providing more migration chances to the users with worse BS associations (e.g., cell-edge users) to improve their performance.

\vspace{-0.05in}
\section{Conclusions}
In this paper, we study the mobility management problem in the multi-cell MEC network, with the goal of minimizing user's energy consumption subject to the reliability constraint for computation offloading. We propose a two-timescale approach with joint optimization of service migration and transmit power control, which is a low-complexity online algorithm and can achieve asymptotical optimality shown by the theoretical analysis. In our approach, the optimal power control for task offloading and the optimal migration policy for BS association, both follow a threshold-based structure. The former uses the threshold to make a binary offloading decision, while the latter uses it to decide whether to migrate from the current BS to the target. These two thresholds are dynamically adjusted to balance the energy and reliability performance. We also extend our two-timescale approach to multiuser management by designing a load-aware multiuser migration scheme. Simulation results demonstrate the superior performance achieved by our approach, especially when the reliability requirement is stringent. For future investigation, we intend to consider a general case that the short-term mobility information is available to be leveraged, which is expected to achieve more proactive migrations. Another direction is considering the cooperative computing among BSs, as an alternative approach to service migration, to deal with user mobility.

\section*{Appendix}

\appendices
\vspace{-0.1in}
\subsection{Proof of Lemma \ref{lem1}}
According to the queue dynamics \eqref{eqn:virtualqueue}, we have
\begin{align}\label{eqn:profA1}
Q(t+1)^2-Q(t)^2&=\left(\max\{Q(t)+X(t)-\epsilon, 0\}\right)^2- Q(t)^2\nonumber \\
&\overset{(a)}{\leq} \left[Q(t)+X(t)-\epsilon\right]^2 -Q(t)^2  \nonumber\\
&\leq X(t)^2+\epsilon^2 +2Q(t)\left[X(t)-\epsilon\right],
\end{align}
where (a) is  derived by $\max\{x, 0\}^2\leq x^2$. Summing the above \eqref{eqn:profA1} over $t\in \{kT, ..., (k+1)T-1\}$ and taking conditional expectation given $Q(kT)$, it follows that $\Delta_T(Q(kT))$ defined in \eqref{eqn:deltaT} is upper bounded by
\begin{align}\label{eqn:profA2}
\Delta_T(Q(kT))&=\mathbb{E}\bigg\{\frac{1}{2}Q((k+1)T)^2 - \frac{1}{2}Q(kT)^2\Big|Q(kT)\bigg\}\nonumber\\
&\leq \mathbb{E}\bigg\{\frac{1}{2}\sum_{t\in\mathcal{T}_k}X(t)^2+\frac{1}{2}\epsilon^2T+\sum_{t\in\mathcal{T}_k}Q(t)\left[X(t)-\epsilon\right]\Big|Q(kT)\bigg\}\nonumber\\
&\overset{(b)}{\leq}\frac{1}{2}\mathbb{E}\bigg\{\sum_{t\in\mathcal{T}_k}a(t)\bigg\}+\frac{1}{2}\epsilon^2T+ \mathbb{E}\bigg\{\sum_{t\in\mathcal{T}_k}Q(t)\left[X(t)-\epsilon\right] \Big|Q(kT)\bigg\}\nonumber\\
&\overset{(c)}{=}\frac{1}{2}(\rho+\epsilon^2)T+\mathbb{E}\bigg\{\sum_{t\in\mathcal{T}_k}Q(t)\left[X(t)-\epsilon\right] \Big|Q(kT)\bigg\},
\end{align}
where $(b)$ is using the facts that $X(t)^2=X(t)\leq a(t)$ and $a(t)$ is independent of $Q(kT)$. \makebox{Step (c)} is because $a(t)$ is i.i.d. over slots with $\mathbb{E}\{a(t)\}=\rho$. Finally, letting $B_1=(\rho+\epsilon^2)/2$ and adding $V\mathbb{E}\left\{\sum_{t\in \mathcal{T}_k}\mathcal{E}(t)|Q(kT)\right\}$ into both sides of \eqref{eqn:profA2} yield the result \eqref{eqn:lemma1}.

\vspace{-0.15in}
\subsection{Proof of Lemma \ref{lem2}}

Since $X(t)\in\{0,1\}$, the queue length $Q(t)$ for each slot $t\in\mathcal{T}_k$ is bounded by
\begin{align}\label{eqn:profB1}
Q(kT)-(t-kT)\epsilon \leq  Q(t) \leq Q(kT) + (t-kT)(1-\epsilon).
\end{align}
Using \eqref{eqn:profB1}, it can be shown that the term $\sum_{t\in\mathcal{T}_k}Q(t)[X(t)-\epsilon]$ in \eqref{eqn:lemma1} can be bounded as
\begin{align}\label{eqn:profB2}
\sum_{t\in\mathcal{T}_k}Q(t)[X(t)-\epsilon] &= \sum_{t\in\mathcal{T}_k}Q(t)X(t) -  \sum_{t\in\mathcal{T}_k}\epsilon Q(t) \nonumber\\
&\leq \sum_{t\in\mathcal{T}_k}\left[Q(kT)+(t-kT)(1-\epsilon)\right]X(t)- \sum_{t\in\mathcal{T}_k}\epsilon \left[Q(kT)-(t-kT)\epsilon\right]\nonumber \\
&\leq \sum_{t\in\mathcal{T}_k}Q(kT)\left[X(t)-\epsilon\right] + \sum_{t\in\mathcal{T}_k}(t-kT)\left[(1-\epsilon)X(t)+\epsilon^2\right].
\end{align}
Taking the conditional expectation on \eqref{eqn:profB2} under a given $Q(kT)$, we have
\begin{align}\label{eqn:profB3}
&\mathbb{E}\bigg\{\sum_{t\in\mathcal{T}_k}Q(t)\left[X(t)-\epsilon\right] \Big|Q(kT)\bigg\}\nonumber \\
\leq~&\mathbb{E}\bigg\{\sum_{t\in\mathcal{T}_k}Q(kT)\left[X(t)-\epsilon\right] \Big|Q(kT)\bigg\}+\mathbb{E}\bigg\{\sum_{t\in\mathcal{T}_k}(t-kT)\left[(1-\epsilon)a(t)+\epsilon^2\right]\bigg\}\nonumber\\
\leq~&\mathbb{E}\bigg\{\sum_{t\in\mathcal{T}_k}Q(kT)\left[X(t)-\epsilon\right] \Big|Q(kT)\bigg\}+ \frac{T(T-1)\left[(1-\epsilon)\rho+\epsilon^2\right]}{2}.
\end{align}
Using the result of \eqref{eqn:profB3} and letting $B_2=B_1+(T-1)\left[(1-\epsilon)\rho+\epsilon^2\right]/2$, we can further relax the inequality \eqref{eqn:lemma1} into \eqref{eqn:lemma2}, which completes the proof.

\vspace{-0.15in}
\subsection{Proof of \eqref{eqn:Znk_sum}}
Since the random variables  $a(t)$ and $h_n(t)$ are i.i.d. over slots $t\in \mathcal{T}_k$, we have
\begin{align}\label{eqn:Znk_sum2}
Z_n^{\emph{sum}}(k)&=\min_{\substack{0\leq p(t)\leq \overline{P}\\[0.1mm] \forall t\in \mathcal{T}_k}} ~\sum_{t\in \mathcal{T}_k}\mathbb{E}\bigg\{V\mathcal{E}(t)+Q(kT)X(t)\Big|n(k)=n\bigg\} \nonumber \\
&=\sum_{t\in \mathcal{T}_k^c}\mathbb{E}\left\{V\cdot0+Q(kT)\cdot\mathds{1}_{\left\{a(t)=1\right\}}\right\}+\sum_{t\in \mathcal{T}_k\backslash\mathcal{T}_k^c}\mathbb{E}\left\{z_n(t)\cdot\mathds{1}_{\left\{a(t)=1\right\}}\right\}\nonumber \\
&=\sum_{t\in\mathcal{T}_k^c} \rho Q(kT)+ \sum_{t\in \mathcal{T}_k\backslash\mathcal{T}_k^c} \rho Z_n(k),
\end{align}
where the second term in the last equality is derived according to the mutual independence between channel gain $h_n(t)$ and task arrival $a(t)$. Based on the definition of set $\mathcal{T}_k^c$,  $Z_n^{\emph{sum}}(k)$ in \eqref{eqn:Znk_sum2} can be further expanded to \eqref{eqn:Znk_sum} for different cases of $n$, which completes the proof.

\vspace{-0.15in}
\subsection{Proof of Proposition \ref{pro3}}
Since $h_n(t)\sim\exp(1/H_n(k))$, i.e., exponential distribution, with some manipulations, $Z_n(k)$ for the channel model \eqref{eqn:hnt} can be expressed as
\begin{align}
Z_n(k)&=\frac{Ve_n(k)}{H_n(k)}\, \mathrm{E}_1\!\left(\frac{h_n^{\rm{min}}(k)}{H_n(k)}\right) +Q(kT)\left(1-e^{-\frac{h_n^{\rm{min}}(k)}{H_n(k)}}\right), \label{eqn:Znk_h}
\end{align}
where $\mathrm{E}_1(x)\triangleq \int_{x}^\infty\frac{e^{-t}}{t}dt$, with $x>0$, is the exponential integral function.

With $f_n(k)=f(k)>\frac{\xi}{\tau_d}$, $\forall n$, it follows that  $\{e_n(k), h_n^{\rm{min}}(k)\}=\{e(k), h^{\rm{min}}(k)\}$, $\forall n$, and $Z_n(k)$ in \eqref{eqn:Znk_h} can be simplified as a function of $H_n(k)$:
\begin{align}\label{eqn:Uk}
Z_n(k)=\frac{Ve(k)}{H_n(k)}\, \mathrm{E}_1\!\left(\frac{h^{\rm{min}}(k)}{H_n(k)}\right) +Q(kT)\left(1-e^{-\frac{h^{\rm{min}}(k)}{H_n(k)}}\right)\triangleq U(H_n(k)), \quad \forall n\in\mathcal{N}.
\end{align}

\subsubsection{Proof of Property a)} We need the following preliminary lemma to prove  Property a):
\begin{lemma}\label{lem3}
$U(H_n(k))$ is a monotonically decreasing function of $H_n(k)>0$.
\end{lemma}
\begin{IEEEproof}
Using the fact that $\left[\int_{f(x)}^{a}g(t)dt\right]^\prime=-g(f(x))\cdot f^\prime(x)$, we have
\begin{align}\label{eqn:Uk_d}
U^\prime(H_n(k))=&-\frac{Ve(k)}{(H_n(k))^2}\, \mathrm{E}_1\!\left(\frac{h^{\rm{min}}(k)}{H_n(k)}\right)  + \frac{Ve(k)}{H_n(k)}\left(-\frac{e^{-\frac{h^{\rm{min}}(k)}{H_n(k)}}}{\frac{h^{\rm{min}}(k)}{H_n(k)}}\right)\cdot -\frac{h^{\rm{min}}(k)}{(H_n(k))^2}\nonumber \\
& -Q(kT) e^{-\frac{h^{\rm{min}}(k)}{H_n(k)}}\cdot \frac{h^{\rm{min}}(k)}{(H_n(k))^2} \nonumber \\
=&-\frac{Ve(k)}{(H_n(k))^2}\left[\mathrm{E}_1\!\left(\frac{h^{\rm{min}}(k)}{H_n(k)}\right)-e^{-\frac{h^{\rm{min}}(k)}{H_n(k)}}\right]- \frac{Q(kT)h^{\rm{min}}(k)}{(H_n(k))^2}e^{-\frac{h^{\rm{min}}(k)}{H_n(k)}}.
\end{align}

According to the definitions of $e_n(k)$ and $h_n^{\rm{min}}(k)$ in Theorem \ref{pro2}, it follows
\begin{align}\label{eqn:veh}
\frac{Ve(k)}{h_n^{\rm{min}}(k)}=\min\left\{Q(kT), V\overline{P}\left[\tau_d-\tfrac{\xi }{f_n(k)}\right]^+\right\}\leq Q(kT).
\end{align}
Therefore, $Ve(k)\leq Q(kT)h^{\rm{min}}(k)$ holds. Plugging it into \eqref{eqn:Uk_d} yields
\begin{align}
U^\prime(H_n(k))\leq -\frac{Q(kT)h^{\rm{min}}(k)}{(H_n(k))^2}\,\mathrm{E}_1\!\left(\frac{h^{\rm{min}}(k)}{H_n(k)}\right) \leq 0,
\end{align}
which completes the proof of Lemma \ref{lem3}.
\end{IEEEproof}

Since $Z_n(k)=U(H_n(k))$,  $\forall n\in \mathcal{N}$, and $U(H_n(k))$ monotonically decreases with $H_n(k)$ by Lemma \ref{lem3}, we have $n^\prime \triangleq \arg\max_{n\in\mathcal{N}\backslash n(k-1)}\left\{Z_n(k)\right\} =\arg\min_{n\in\mathcal{N}\backslash n(k-1)}\left\{H_n(k)\right\}$, completing the proof.

\subsubsection{Proof of Property b)} Recall that in the migration policy \eqref{eqn:associa_policy}, the user migrates from its current associated BS $n(k-1)$ to BS $n^\prime$ if $(1-\alpha)Z_{n^\prime}(k)+\alpha Q(kT)<Z_{n(k-1)}(k)$ is met. Plugging \eqref{eqn:Uk} into the condition and simplifying, we obtain
\begin{align}\label{eqn:associa_policy_b}
(1-\alpha)\left[\frac{Ve(k)}{H_{n^\prime}(k)}\, \mathrm{E}_1\left(\frac{h^{\rm{min}}(k)}{H_{n^\prime}(k)}\!\right) -Q(kT)e^{-\frac{h^{\rm{min}}(k)}{H_{n^\prime}(k)}}\right]& \nonumber \\
<\frac{Ve(k)}{H_{n(k-1)}(k)}\, &\mathrm{E}_1\!\left(\frac{h^{\rm{min}}(k)}{H_{n(k-1)}\!(k)}\!\right) -Q(kT)e^{-\frac{h^{\rm{min}}(k)}{H_{n(k-1)}\!(k)}}.
\end{align}

Note that for $x>0$, $\frac{1}{2}e^{-x}\ln(1+\frac{2}{x})<\mathrm{E}_1(x)<e^{-x}\ln(1+\frac{1}{x})$. Utilizing this property, \eqref{eqn:associa_policy_b} can be re-written as the following sufficient condition:
\begin{align}\label{eqn:associa_policy_c}
\underbrace{\vphantom{\left(\frac{H_{n^\prime}(k)}{h^{\rm{min}}(k)}\right)} e^{-\frac{h^{\rm{min}}(k)}{H_{n^\prime}(k)}+\ln(1-\alpha)}}_{A_1}&\underbrace{\left[\frac{Ve(k)}{H_{n^\prime}(k)}\ln\left(1+\frac{H_{n^\prime}(k)}{h^{\rm{min}}(k)}\right)-Q(kT)\right]}_{B1}\nonumber\\
&\qquad \qquad <\underbrace{\vphantom{\left(\frac{H_{n^\prime}(k)}{h^{\rm{min}}(k)}\right)}e^{-\frac{h^{\rm{min}}(k)}{H_{n(k-1)}\!(k)}}}_{A_2}\underbrace{\left[\frac{Ve(k)}{2H_{n(k-1)}\!(k)}\ln\left(1+\frac{2H_{n(k-1)}\!(k)}{h^{\rm{min}}(k)}\right)-Q(kT)\right]}_{B_2}.
\end{align}
Here, $\frac{Ve(k)}{H_{n^\prime}(k)}\ln\left(1+\frac{H_{n^\prime}(k)}{h^{\rm{min}}(k)}\right)< \frac{Ve(k)}{H_{n^\prime}(k)}\frac{H_{n^\prime}(k)}{h^{\rm{min}}(k)}=\frac{Ve(k)}{h^{\rm{min}}(k)}\leq Q(kT)$, where the last equality is according to \eqref{eqn:veh}. Similarly, $\frac{Ve(k)}{2H_{n(k-1)}\!(k)}\ln\left(1+\frac{2H_{n(k-1)}\!(k)}{h^{\rm{min}}(k)}\right)< Q(kT)$. Thus, we have $B_1, B_2<0$. With $A_1, A_2>0$, \eqref{eqn:associa_policy_c} can be written as $A_1\cdot(-B_1)>A_2\cdot(-B_2)$, and decomposed into the following two conditions by letting $A_1>A_2$ and $-B_1>-B_2$, respectively:
\begin{gather}
-\frac{h^{\rm{min}}(k)}{H_{n^\prime}(k)}+\ln(1-\alpha)> -\frac{h^{\rm{min}}(k)}{H_{n(k-1)}\!(k)}, \label{eqn:associa_policy_d} \\
\frac{1}{H_{n^\prime}(k)}\ln\left(1+\frac{H_{n^\prime}(k)}{h^{\rm{min}}(k)}\right)<\frac{1}{2H_{n(k-1)}\!(k)}\ln\left(1+\frac{2H_{n(k-1)}\!(k)}{h^{\rm{min}}(k)}\right).\label{eqn:associa_policy_e}
\end{gather}
It can be checked that $\frac{1}{x}\ln\left(1+x\right)$ is monotonically decreasing with $x$; thus \eqref{eqn:associa_policy_e} is equivalent to $H_{n^\prime}(k)>2H_{n(k-1)}(k)$. By letting $H_\text{th}=\frac{h^{\rm{min}}(k)}{\ln(\frac{1}{1-a})}$,  \eqref{eqn:associa_policy_d} is equivalent to $H_{n^\prime}(k)>\frac{H_\text{th}(k)H_{n(k-1)}(k)}{H_\text{th}(k)-H_{n(k-1)}(k)}$, with $H_{n(k-1)}(k)<H_\text{th}(k)$. Summarizing above conditions yields the results of Property b).

\vspace{-0.15in}
\subsection{Proof of Proposition \ref{pro4}}
For notational simplicity, we use variable $\nu_n(k)$ to replace $\frac{h_n^{\rm{min}}(k)}{H_n(k)}$ for all $n\in\mathcal{N}$ in this proof.

Recall that $Z_n(k)$ for the channel model \eqref{eqn:hnt} is given by \eqref{eqn:Znk_h}. By plugging $\nu_n(k)=\frac{h_n^{\rm{min}}(k)}{H_n(k)}$ into \eqref{eqn:Znk_h}, we have
\begin{align}
Z_n(k)&=\frac{Ve_n(k)}{h_n^{\rm{min}}(k)} \,\nu_n(k) \, \mathrm{E}_1\!\left(\nu_n(k)\right) +Q(kT)\left(1-e^{-\nu_n(k)}\right). \label{eqn:Znk_h2}
\end{align}

Assume that $\overline{P}>\frac{Q}{V\left[\tau_d-\frac{\xi}{f_n(k)}\right]^+}$, for all $n$. Then, according to \eqref{eqn:veh}, $\frac{Ve_n(k)}{h_n^{\rm{min}}(k)}=Q(kT)$ and $Z_n(k)$ can be further expressed as
\begin{align}\label{eqn:Znk_h3}
Z_n(k)&=Q(kT)+Q(kT)\left[\nu_n(k) \, \mathrm{E}_1\!\left(\nu_n(k)\right) - e^{-\nu_n(k)}\right] \triangleq Y(\nu_n(k)).
\end{align}

\subsubsection{Proof of Property a)} Similar to the proof of Proposition \ref{pro3}, $ n^\prime = \argmin_{n\in\mathcal{N}\backslash n(k-1)} \left\{\nu_n(k)\right\}$ is sufficient to verifying $Y(\nu_n(k))$ is a monotonic increasing function with $\nu_n(k)$. Since $\mathrm{E}_1^\prime(x)=-e^{x}/x$, we have
\begin{align}
Y^\prime(\nu_n(k))= Q(kT) \left[\mathrm{E}_1\!\left(\nu_n(k)\right)+\nu_n(k)\frac{-e^{\nu_n(k)}}{\nu_n(k)}+e^{\nu_n(k)}\right]=Q(kT)\,\mathrm{E}_1\!\left(\nu_n(k)\right)
\end{align}
Since $Z_n(k) = Y(\nu_n(k))$, $\forall n\in \mathcal{N}$, and $Y(\nu_n(k))$ is monotonically increasing with $\nu_n(k)$, we have $n^\prime = \argmin_{n\in\mathcal{N}\backslash n(k-1)}\{Z_n(k)\}=\argmin_{n\in\mathcal{N}\backslash n(k-1)}\{\nu_n(k)\}$, which completes the proof.

\subsubsection{Proof of Property b)} Letting $(1-\alpha)Z_{n^\prime}(k)+\alpha Q(kT)<Z_{n(k-1)}(k)$ and simplifying, we have the migration condition for this case:
\begin{align}
(1-\alpha)\Big[\nu_{n^\prime}(k)\, \mathrm{E}_1\left(\nu_{n^\prime}(k)\right)-e^{-\nu_{n^\prime}(k)}\Big]<\nu_{n(k-1)}(k)\, \mathrm{E}_1\left(\nu_{n(k-1)}(k)\right)-e^{-\nu_{n(k-1)}(k)}
\end{align}
Similar to the proof in Proposition \ref{pro3}, using the facts that $\frac{1}{2}e^{-x}\ln(1+\frac{2}{x})<\mathrm{E}_1(x)<e^{-x}\ln(1+\frac{1}{x})$ and $x\ln(1+\frac{1}{x})<1$ hold for $x>0$,  we have
\begin{align}
\underbrace{\vphantom{\left[\ln\left(\tfrac{1}{v_{n^\prime}(k)}\right)\right]}
e^{-\nu_{n^\prime}(k)+\ln(1-\alpha)}}_{C_1}\underbrace{\left[1-\nu_{n^\prime}(k)\ln\!\left(1+\tfrac{1}{\nu_{n^\prime}(k)}\right)\right]}_{D_1}>
\underbrace{\vphantom{\left[\ln\left(\tfrac{1}{v_{n^\prime}(k)}\right)\right]}
e^{-\nu_{n(k-1)}\!(k)}}_{C_2}\underbrace{\left[1-\tfrac{\nu_{n(k-1)}\!(k)}{2}\ln\!\left(1+\tfrac{2}{\nu_{n(k-1)}\!(k)}\right)\right]}_{D_2}
\end{align}
where components $C_1, C_2, D_1, D_2$ are all positive. Letting $C_1>C_2$ and $D_1>D_2$, it follows
\begin{gather}
\nu_{n^\prime}(k) +\ln\left(\tfrac{1}{1-\alpha}\right)< \nu_{n(k-1)}(k) \label{eqn:associa_policy_h}\\
\nu_{n^\prime}(k)\ln\!\left(1+\tfrac{1}{\nu_{n^\prime}(k)}\right)< \tfrac{\nu_{n(k-1)}\!(k)}{2}\ln\!\left(1+\tfrac{2}{\nu_{n(k-1)}\!(k)}\right) \label{eqn:associa_policy_g}
\end{gather}
Note that $x\ln(1+\frac{1}{x})$ is monotonically increasing with $x$. Thus,  \eqref{eqn:associa_policy_g} is equivalent to $2\nu_{n^\prime}(k)<\nu_{n(k-1)}$. Combining this with \eqref{eqn:associa_policy_h} gives the result of Property b).

\vspace{-0.15in}
\subsection{Proof of Theorem \ref{pro5}}
Using Lemma \ref{lem2} and the fact that the proposed algorithm is developed through minimizing the R.H.S. of the inequality \eqref{eqn:lemma2}, we have
\begin{align}
&\Delta_T(Q^*(kT)) +V\mathbb{E}\bigg\{\sum_{t\in \mathcal{T}_k}\mathcal{E}^*(t)\Big|Q^*(kT)\bigg\}\nonumber \\
\leq~& B_2T +\mathbb{E}\bigg\{\sum_{t\in\mathcal{T}_k}V\mathcal{E}^*(t)+Q^*(kT)\left[X(t)^*-\epsilon\right]\Big|Q^*(kT)\bigg\} \label{eqn:thm2_pf1} \\
\overset{(d)}{\leq}~& B_2T +\mathbb{E}\bigg\{\sum_{t\in\mathcal{T}_k}V\widehat{\mathcal{E}}(t)+Q^*(kT)\left[\widehat{X}(t)-\epsilon\right]\Big|Q^*(kT)\bigg\} \nonumber \\
\overset{(e)}{\leq}~& B_2T +\mathbb{E}\bigg\{\sum_{t\in\mathcal{T}_k}V\widehat{\mathcal{E}}(t)\Big|Q^*(kT)\bigg\}- Q^*(kT) \delta T.
\end{align}
Here, $\widehat{\mathcal{E}}(t)$ and $\widehat{X}(t)$ denote the energy consumption and task failure achieved by the policy satisfying the conditions \eqref{eqn:thm2_condi}, respectively. Step (d) is because the right term of \eqref{eqn:thm2_pf1} obtained by the proposed algorithm is not more than that of any other feasible policy including the policy satisfying the conditions \eqref{eqn:thm2_condi}. Step (e) is derived by the conditions \eqref{eqn:thm2_condi}.

Rearranging the terms and noting that $|\mathcal{E}^*(t)-\widehat{\mathcal{E}}(t)|\leq \overline{P}\cdot\max\limits_{n\in \mathcal{N}}\left\{\tau_d-\frac{\xi}{f_{n}^\text{max}}\right\}\triangleq E_{\rm{max}}$, we have
\begin{align}
\Delta_T(Q^*(kT)) \leq B_2T +VT E_{\rm{max}}- Q^*(kT) \delta T.
\end{align}
Taking expectation of the above and summing it over $k=0, 1, ..., K-1$ yield
\begin{align}
\frac{1}{2}\mathbb{E}\left\{Q^*(kT)^2\right\} -
\frac{1}{2}\mathbb{E}\left\{Q^*(0)^2\right\} \leq K\left[B_2T +VT E_{\rm{max}}\right] -  \delta T \sum_{k=0}^{K-1}\mathbb{E} \left\{Q^*(kT)\right\}.
\end{align}
Rearranging terms in the above, dividing both sides of $K\delta T$, and taking limit as $K \rightarrow\infty$ yield
\begin{align}
\lim_{K \rightarrow\infty} \frac{1}{K}\sum_{k=0}^{K-1}\mathbb{E} \left\{Q^*(kT)\right\} \leq \frac{B_2+VE_{\rm{max}}}{\delta} +\frac{ \mathbb{E}\left\{Q^*(0)^2\right\}-\mathbb{E}\left\{Q^*(kT)^2\right\}}{2 K\delta T} \leq   \frac{B_2+VE_{\rm{max}}}{\delta}
\end{align}
This proves \eqref{eqn:thm21} in Theorem \ref{pro5}.

According to \cite[Theorem 4.5]{Lyapunov1}, if the problem is feasible, there exists a stationary optimal $\omega$-only policy, in which decisions $n(k)$ and $\{p(t)\}$ are made independent of the queue length, achieving the minimum energy consumption $\mathit{E}_{\rm{av}}^\emph{opt}$ while meeting the queue stability constraint. Therefore, we have
\begin{align}\label{eqn:thm2_pf2}
\Delta_T(Q^*(kT)) +V\mathbb{E}\bigg\{\sum_{t\in \mathcal{T}_k}\mathcal{E}^*(t)\Big|Q^*(kT)\bigg\}&\leq B_2T + V \mathbb{E}\bigg\{\sum_{t\in\mathcal{T}_k} \mathcal{E}^\emph{opt}(t) \bigg\}
\end{align}
where the term $Q^*(kT)\mathbb{E}\{\sum_{t\in\mathcal{T}_k} [X^\emph{opt}(t)-\epsilon]\}$ is neglected in the R.H.S. of \eqref{eqn:thm2_pf2} since it is non-positive due to satisfying the queue stability constraint.

Taking expectations of the above inequality and summing it over $k=0, 1, ..., K-1$ yield
\begin{align}
&\frac{1}{2}\mathbb{E}\left\{Q^*(kT)^2\right\} -
\frac{1}{2}\mathbb{E}\left\{Q^*(0)^2\right\}+V\sum_{k=0}^{K-1}\mathbb{E}\bigg\{\sum_{t\in \mathcal{T}_k}\mathcal{E}^*(t)\bigg\} \nonumber \\
&\qquad \qquad \qquad \qquad \qquad \qquad  \qquad \qquad \qquad \qquad  \leq  KB_2T + V\sum_{k=0}^{K-1}\mathbb{E}\bigg\{\sum_{t\in\mathcal{T}_k}\mathcal{E}^\emph{opt}(t) \bigg\}.
\end{align}
Dividing both sides by $VKT$, taking the limit as $K \rightarrow\infty$, and noting that $\{h_n(t), a(t)\}$ are i.i.d. over slots within a frame, we have
\begin{align}
\lim_{K \rightarrow\infty} \frac{1}{K}\sum_{k=0}^{K-1}\sum_{t\in \mathcal{T}_k} \mathbb{E}\left\{\mathcal{E}^*(t)\right\} \leq \frac{B_2}{V} + \lim_{K \rightarrow\infty} \frac{1}{K}\sum_{k=0}^{K-1}\sum_{t\in \mathcal{T}_k}\mathbb{E}\left\{ \mathcal{E}^\emph{opt}(t)\right\} =  \frac{B_2}{V} +\mathit{E}_{\rm{av}}^\emph{opt}.
\end{align}
This yields \eqref{eqn:thm22} in Theorem \ref{pro5}.

\vspace{-0.1in}
\bibliographystyle{IEEEtran}
\bibliography{IEEEabrv,link}

\end{document}